\documentclass[aps,prb,amsmath,superscriptaddress,amssymb,longbibliography,twocolumn]{revtex4-2}
\usepackage{graphicx} 
\usepackage{physics}  
\usepackage{braket}   
\usepackage{comment}  
\usepackage{color}  
\usepackage[normalem]{ulem} 
\usepackage[colorlinks=true,citecolor=blue,linkcolor=magenta,breaklinks=true]{hyperref} 

\begin{document}

\title{Long-range model for electronic pseudospin-nuclear-spin coupling and two-stage ordering in TmVO$_4$}
\date{\today}
\begin{abstract}
We study an infinite-range coupled electronic-pseudospin and nuclear-spin model for ferroquadrupolar and nuclear-spin ordering in TmVO$_4$ in external magnetic and strain fields. This material is an experimental realization of a transverse-field Ising model, where the Ising degree of freedom is quadrupolar and non-magnetic, but a transverse component is magnetic and couples both to external magnetic fields and to the nuclear spins via a hyperfine coupling. At zero external magnetic field, the electronic and nuclear degrees of freedom undergo well-separated two-stage ordering, accompanied by the release of their respective entropies. A transverse magnetic field polarizes the electronic orbital moments and, via the hyperfine coupling, the nuclear spins. The quadrupolar ordering temperature is gradually reduced to zero. However, the nuclear transition is replaced by a crossover at nonzero magnetic field. Quantum fluctuations are enhanced near the phase transitions and lead to peaks in the magnetic susceptibility. The spectral functions reveal a softening of a low-energy mode near the quantum critical point, consistent with the closing of the excitation gap and its reopening in the disordered phase. These features provide direct dynamical signatures of the field-driven quantum critical phenomena.
\end{abstract}

\author{Sayan Ghosh}
\affiliation{S.N. Bose National Centre for Basic Sciences, Kolkata 700098, India.}
\author{Anirudha Menon}
\affiliation{Department of Physics, University of California Davis, Davis, California 95616, USA}
\affiliation{Centre for Theoretical and Computational Physics, National Yang Ming Chiao Tung University, Hsinchu City, Taiwan}
\author{Manoranjan Kumar}
\email{manoranjan.kumar@bose.res.in}
\affiliation{S.N. Bose National Centre for Basic Sciences, Kolkata 700098, India.}
\author{Rajiv R. P. Singh}
\email{rrpsing@ucdavis.edu}
\thanks{The last two authors contributed equally to this work.}
\affiliation{Department of Physics, University of California Davis, Davis, California 95616, USA}

\maketitle
\section{Introduction} 

Rare-earth compounds exhibit many unconventional magnetic phases, such as quantum spin liquids (QSLs) in YbMgGaO$_4$
\cite{xu2016absence,li2016muon,paddison2017continuous,li2015rare},
Tb$_2$Ti$_2$O$_7$
\cite{kadowaki2019spin,wakita2016quantum,kadowaki2018continuum,takatsu2011quantum}, and $\alpha$-RuCl$_3$ \cite{baek2017evidence,zheng2017gapless,wang2016spin,banerjee2017neutron}, spin ice in Dy$_2$Ti$_2$O$_7$ \cite{kassner2015supercooled,mcclarty2015chain,ross2011quantum}
and Ho$_2$Ti$_2$O$_7$
\cite{ross2011quantum,clark2021quantum},
and multipolar order (octupolar, quadrupolar) in CeB$_6$, NpO$_2$,
and PrTi$_2$Al$_{20}$
\cite{thalmeier1998theory,shiina2003dynamics,thalmeier2004multipolar,
sakai2004invariant,caciuffo2003multipolar,suzuki2010first,
tokunaga2006nmr,sakai2025interplay,sato2012ferroquadrupolar,
taniguchi2019field}. These diverse phases arise from the highly localized electrons in partially filled $4f$ shells, which are shielded by the outer $5s$ and $5p$ orbitals and subject to strong spin-orbit coupling
\cite{jensen1991rare,hughes2007lanthanide}. Among these, the tetragonal insulator TmVO$_4$ exhibits canonical ferroquadrupolar (FQ) order. This quadrupolar order manifests as electronic nematicity through the spontaneous ordering of quadrupolar moments associated with the electronic orbitals \cite{maharaj2017transverse, rosenberg2019divergence}. The ferroquadrupolar phase is a symmetry-broken state characterized by the uniform alignment of electronic quadrupole moments, which produces a spontaneous lattice distortion. Unlike conventional magnetic dipole order, ferroquadrupolar (FQ) order arises from interactions between higher-order multipole moments-specifically, the anisotropic distribution of electronic charge associated with orbital degrees of freedom \cite{wang2016spin,sato2012ferroquadrupolar,ishii2019anisotropic,massat2022field,akazawa1998quadrupolar}. In this material, the Tm$^{3+}$ ions ($4f^{12}$) possess a low-energy non-Kramers doublet associated with their localized $4f$ electronic degrees of freedom \cite{knoll1971absorption,zic_thesis}. These localized $4f$ degrees of freedom give rise to the relevant quadrupolar degrees of freedom and are distinct from the nuclear-spin degrees of freedom. These doublets order through a cooperative Jahn–Teller effect into the FQ phase below $T_Q\approx2.14\,$K, and break $D_{4h}$ symmetry down to $D_{2h}$ through a $B_{2g}$ shear strain~\cite{segmuller1974x,gehring1973cooperative,massat2022field}.

Experimental studies of TmVO$_4$ suggest a rich phase diagram under strain, magnetic fields, and chemical substitution \cite{massat2022field,vinograd2022second,nian2024spin,wang2021anisotropic,zic2024realization,nian2023spin}. The transition of the ferroquadrupolar (FQ) phase to the paraquadrupolar (PQ) phase of this material can be tuned to a quantum critical point using either a magnetic field along the c-axis or a $B_{1g}$ strain field, both of which act as transverse fields on the pseudospin doublet degree of freedom. At low magnetic fields, the phase boundary follows a semiclassical mean-field description of the transverse-field Ising model (TFIM). However, at higher fields, quantum fluctuations cause deviations from this behavior, highlighting the role of the crystal lattice in mediating interactions between local quadrupoles and influencing the critical scaling exponents \cite{massat2022field}. Elastoresistivity measurements and strain-tuned calorimetry reveal
classical critical behavior at $T_Q$ associated with the
Ising-symmetric ferroquadrupolar (electronic-nematic) order parameter,
while NMR studies demonstrate quantum critical fluctuations within a
fan-like region in the temperature--transverse magnetic-field plane
\cite{massat2022field,vinograd2022second,nian2024spin,zic2024realization}.
Theoretical analyses based on an infinite‐range transverse-field Ising model capture many aspects of the quantum critical phenomena~\cite{curro2024quantum,nian2024spin}.

Recently, attention has turned to the phase diagram at very low temperatures and the feedback of the nuclear-spin degrees of freedom on the electronic phase diagram \cite{zic_thesis,zic2024realization,zic2025electro}. Experimentally, the phase boundary was identified from a drop in the magnetic susceptibility \cite{zic_thesis,zic2024realization,zic2025electro}, consistent with the discontinuous change predicted by mean-field theory. At low temperatures, the phase boundary bends back toward lower fields. This is in contrast to the well-known TFIM material LiHoF$_4$ \cite{Bitko1996}, where the phase boundary bends towards higher fields due to the coupling to the nuclear degrees of freedom. The difference arises because the hyperfine coupling in LiHoF$_4$ is longitudinal to the Ising degree of freedom whereas it is transverse in TmVO$_4$. Within a single-site mean-field theory for a TFIM coupled to a nuclear spin via a hyperfine coupling, Zic et al. \cite{zic2024realization,zic2025electro,zic_thesis} have shown that the bending of the phase boundary is in agreement with the model. At zero transverse field, mean-field theory predicts spontaneous ordering of the nuclear spins at a temperature scale set by the square of the hyperfine coupling. Such ordering has not yet been observed in the experiments down to much lower temperatures than predicted by mean-field theory \cite{zic2024realization,Suzuki1980}.

It is well-known that the mean-field theory lacks quantum critical phenomena as the gap remains finite at the transition and no entanglement develops near the quantum critical point. In this work, we address these limitations by introducing a long-range Ising pseudospin–$1/2$ model \cite{curro2024quantum} that includes (i) ferroquadrupolar Jahn–Teller interactions, (ii) a transverse $B_{1g}$ strain field $h_y$, (iii) a transverse c-axis magnetic field $h_x$, and (iv) a hyperfine coupling to nuclear spins of strength $A$. We use exact diagonalization and mean-field theory to solve the model and first benchmark the pure electronic limit ($A=0$) against single-site mean-field theory and recent experiments, recovering the phase diagram and emergent rotational symmetry in the transverse-field $(h_x,h_y)$ space. 

We then include the hyperfine coupling to reveal a two-step ordering sequence: a sharp electronic transition at $T_Q$ (entropy reduction from $\ln4$ to $\ln2$) followed, under strain but not under transverse magnetic field, by a genuine nuclear transition at $T_N\sim A^2/J_z$ (entropy goes from $\ln2\to0$) where $J_z$ is the ferroquadrupolar stiffness. In the transverse magnetic field case, the model confirms the weak back-bending of the electronic quadrupolar phase diagram near the $T=0$ transition while the lower‐$T$ nuclear transition is replaced by a smooth crossover of the nuclear polarization. We show that the energy gap goes to zero as the quantum critical point is approached from both the paraquadrupolar and ferroquadrupolar sides. Away from the quantum critical point, the spectral weight is dominated by a single gapped mode. However, at the quantum critical point, several states approach zero energy and carry significant spectral weight. 
Finally, we investigate the strain dependence of the entropy and discuss its implications for possible elastocaloric effects under adiabatic strain sweeps at cryogenic temperatures.

The remainder of this paper is organized as follows. In Sec.~II, we introduce the effective infinite-range pseudospin Hamiltonian for
TmVO$_4$, including the mapping of the low-energy non-Kramers doublet
onto pseudospin-$1/2$ operators, the coupling to strain and magnetic
fields, and the hyperfine interaction with the nuclear spins. We also
describe the symmetry reduction used in the exact-diagonalization
calculations and define the thermodynamic and dynamical observables.
In Sec.~III, we benchmark the model in the absence of nuclear coupling
and determine the ferroquadrupolar phase boundary and transverse
susceptibility. Sec.~IV discusses the effects of the hyperfine
interaction, including the two-stage ordering, field-induced nuclear
crossover, excitation gaps, and dynamical structure factors. The strain
dependence of the entropy and nuclear ordering scale is discussed in
Sec.~V. The appendices provide additional details on the combinatorial
factors, adaptive nuclear-sector selection, mean-field analysis, and
finite-size scaling.



\section{Model Hamiltonian}

The localized $4f$ electronic degrees of freedom of the Tm$^{3+}$
ions experience a highly anisotropic crystal-field environment,
giving rise to Ising-type interactions within the low-energy
non-Kramers doublet. We refer to these low-energy $4f$ degrees of
freedom as the electronic sector, represented by an effective
pseudospin-$1/2$ operator $\mathbf{S}_i$. No additional
itinerant-electron or charge degrees of freedom are included in the
model. In TmVO$_4$, the ferroquadrupolar interaction is mediated by lattice
strain. Following refs.~\cite{massat2022field,zic_thesis,zic2025electro}, we consider the coupling between the quadrupolar moment $P_{xy}$ and the $B_{2g}$ strain field,
\begin{equation}
H_s =
\sum_i V_s \epsilon_{B_{2g}} P_{xy}(i)
+\frac{C_{66}}{2}\epsilon_{B_{2g}}^2,
\end{equation}
where
\begin{equation}
P_{xy} = J_xJ_y+J_yJ_x.
\end{equation}
For a uniform static strain, minimizing $H_s$ with respect to
$\epsilon_{B_{2g}}$ gives
\begin{equation}
H_s^{\rm eff}
=-\frac{V_s^2}{2C_{66}}
\sum_{i,j}P_{xy}(i)P_{xy}(j).
\end{equation}
Thus, the coupling to the elastic degrees of freedom generates an
infinite-range ferroquadrupolar interaction. The lowest $E_g$ crystal-field doublet can be represented by an effective pseudospin-$1/2$. Within this doublet, the relevant operators are mapped onto pseudospin operators as
\begin{equation}
P_{xy}\rightarrow S^z,\qquad
O_2^2=J_x^2-J_y^2\rightarrow S^y,\qquad
J_z\rightarrow S^x.
\end{equation}
The corresponding conjugate fields are the $B_{2g}$ strain $\epsilon_{B_{2g}}$, the $B_{1g}$ strain $\epsilon_{B_{1g}}$, and the magnetic field $H_z$, respectively. Consequently, the strain field and the magnetic field enter the effective pseudospin Hamiltonian as transverse fields acting on different pseudospin components. We therefore identify the ferroquadrupolar coupling as
\begin{equation}
J_z=\frac{V_s^2}{2C_{66}},
\end{equation}
while \(h_x\) and \(h_y\) denote the effective pseudospin fields associated with the applied magnetic field \(H_z\) and the \(B_{1g}\) strain \(\epsilon_{B_{1g}}\), respectively. For a system
of $N$ Tm sites, the resulting Hamiltonian including the hyperfine
coupling to the nuclear spins is 
\begin{equation}\label{eq:H}
\begin{split}
H=-\frac{2J_z}{N}\Bigl(\sum_{i}S_{i}^{z}\Bigr)^{2}
+2h_{y}\sum_{i}S_{i}^{y}
+2h_{x}\sum_{i}S_{i}^{x}\\
+4A\sum_{i}I_{i}^{z}(\frac{1}{N}\sum_jS_{j}^{x}),
\end{split}
\end{equation}
where $\vec{S}_i$ are pseudospin-1/2 operators for the non-Kramers doublet at site $i$. $I_i^z$ is the $z$-component of the nuclear-spin operator, corresponding to the naturally abundant Tm isotope with nuclear spin $I=1/2$, and couples to the electronic doublet through the hyperfine interaction of strength A. $J_{z}$ is the ferroquadrupolar stiffness, which sets the energy scale for the zero-field transition temperature $T_Q$, and $h_{y}$ and $h_{x}$ are strain and transverse fields, respectively. To ensure a well-defined thermodynamic limit, the interaction strength is normalized by a factor of $1/N$, so that the total interaction energy remains extensive while each spin interacts equally with all others. If all electronic and nuclear spins are treated individually, the Hilbert-space dimension of an $N$-site system scales as $4^N$. In this infinite-range model we can write the Hamiltonian in terms of a large total electronic pseudospin $\vec{S}_t=\sum_i \vec{S}_i$.

The local hyperfine interaction has the form $4A\sum_i I_i^zS_i^x$. In the present infinite-range description, we retain only its uniform ($q=0$) component and replace it by the collective coupling $\frac{4A}{N}I_t^zS_t^x$, where $I_t^z=\sum_i I_i^z$ and $S_t^x=\sum_i S_i^x$. This approximation is consistent with the collective nature of the pseudospin Hamiltonian, whose ferroquadrupolar interaction is formulated entirely in terms of the uniform ($q=0$) component of the electronic pseudospin. Since the electronic interaction is represented by an infinite-range coupling, the low-energy physics is governed by the uniform ($q=0$) collective mode, and coupling the nuclei to the same collective mode provides the natural long-wavelength counterpart of the local hyperfine interaction. The collective hyperfine term therefore retains the essential feedback between the uniform nuclear polarization and the electronic transverse magnetization while preserving the permutation symmetry of the infinite-range model. Consequently, it captures the mechanism responsible for the field-induced modification of the electronic phase boundary, the two-stage entropy release, and the crossover of the nuclear polarization in a transverse magnetic field. We also note that NMR spin-echo experiments are done by preparing the nuclei in a uniform (q=0) state, which is another reason for considering this simplified coupling \cite{nian2024spin}.

Since the total magnetization along $z$, $I_t^z$, commutes with the Hamiltonian, the Hamiltonian decomposes into independent nuclear‐spin sectors labeled by $I_t^z=m$, where $m$ takes values from $-N/2,\dots,N/2$.  In each sector, the Hamiltonian is further block diagonal in the total electronic-pseudospin quantum number $s_t=0,\dots,N/2$, yielding blocks of matrix dimension $(2s_t+1)\times(2s_t+1)$.  Each $m$ sector is handled separately: for fixed $m$ we diagonalize the $(2s_t+1)$‐dimensional Hamiltonian matrices for all allowed $s_t$. This approach greatly simplifies the calculation by avoiding having to simultaneously deal with an exponentially large Hilbert space. Thus, one can numerically study systems composed of thousands of spins. Two combinatorial factors arise from the presence of two distinct sectors: $d(m)$ for the nuclear spin sector and $d(s_t)$ for the electronic pseudospin sector. A detailed account of the combinatorial factors is given in Appendix \ref{App:B}. For the system sizes considered in this work, the above procedure can be carried out exactly by summing over all nuclear-spin sectors. However, for the largest system sizes and in the low-temperature regime, as required for determining the finite-temperature phase boundary, the computational cost of treating every nuclear-spin sector becomes prohibitively expensive. 
To overcome this limitation, we employ an adaptive nuclear-spin sector selection scheme. In this approach, we first identify the dominant nuclear-spin sector, $m^*$, which provides the largest contribution to the partition function. We then compute the low-lying eigenvalues and eigenvectors using exact diagonalization only for the nuclear-spin sectors within the window $m^*-\Delta m \le m \le m^*+\Delta m$. These eigenvalues and eigenvectors are subsequently used to construct the partition function and evaluate thermodynamic quantities accurately at low temperatures. The details of this procedure are presented in Appendix~\ref{App:D}.


 The system, in the absence of external fields, is in a FQ ordered state below the critical temperature $T_Q$, and a  paraquadrupolar (PQ) phase above $T_Q$. To locate the electronic phase transition in the symmetry unbroken state of the model, we compute the square of the order-parameter $\langle(S^z)^2\rangle$ and the Binder cumulant $U_{\rm el}$ defined as:  
\begin{equation}
U_{\rm el} = 1 - \frac{\langle (S^{z})^{4}\rangle}{3\langle (S^{z})^{2}\rangle^{2}},
\end{equation}
Here, $\langle \cdots \rangle$ denotes the combined quantum and thermal averages at finite $T$. The expectation value of any operator $\hat O$ for system size $N$ can then be written as 
\begin{equation}\label{eq:op_partitionfunc}
\begin{split}
\langle \hat{O} \rangle = \frac{1}{Z} \sum_{m = -N/2}^{N/2} d(m) \sum_{s_t = 0}^{N/2} d(s_t) \sum_{i} \langle \psi_i | \hat{O} | \psi_i \rangle e^{-E_i(m, s_t)/T},
\\
Z = \sum_{m = -N/2}^{N/2} d(m) \sum_{S_t = 0}^{N/2} d(s_t) \sum_{i} e^{-E_i(m, s_t)/T}.
\end{split}
\end{equation}
The phase boundary can be obtained either by studying the crossings of the Binder cumulants $U_{\rm el}$ for different system sizes or by locating the steepest change or minima in $\frac{d\langle(S^z)^2\rangle}{dh_x}$ vs $h_x$ at different temperatures or by locating the minima in $\frac{d\langle (S^z)^2 \rangle}{dT}$ vs $T$ for different $h_x$. 

At very low temperatures, we found the optimal method to be an adaptive nuclear-sector selection algorithm introduced in Appendix~\ref{App:D} to compute $d\langle (S^z)^2\rangle/dh_x$. The minima of this quantity as a function of $h_x$ for different system sizes are then extrapolated to the thermodynamic limit, yielding the low-temperature phase boundary for finite hyperfine coupling $A$.  

We also calculate the entropy $S = -\left( \frac{\partial F}{\partial T} \right)_V$ and the free energy $F = -k_B T \log_e(Z)$, where the partition function $Z$ is defined  in Eq.~(\ref{eq:op_partitionfunc}) and $k_B$  is the Boltzmann constant.

The spectral function associated with the total transverse spin operator,
$S_{xx}(\omega)$, is computed as
\begin{equation} \label{eq:spectral_func}
S^{\alpha\alpha}(\omega) =
\sum_{n}
\frac{
\left|\langle \psi_n \lvert S^\alpha \rvert \psi_{\mathrm{gs}} \rangle\right|^2
}{
E_n - (E_0+\omega) + i\eta
},
\end{equation}
where $\alpha = x,y$, or $z$ and $\eta = 10^{-3}$ is the Lorentzian broadening factor. Here, $\psi_{\mathrm{gs}}$ and $\psi_n$ denote the ground state and the excited eigenstates with energies $E_0$ and $E_n$, respectively. The sum runs over excited states belonging to the same nuclear-spin sector, characterized by total nuclear spin $I_t^z = m = -N/2$.

\begin{figure}[h]
    \centering
    \includegraphics[width=1.0\columnwidth]{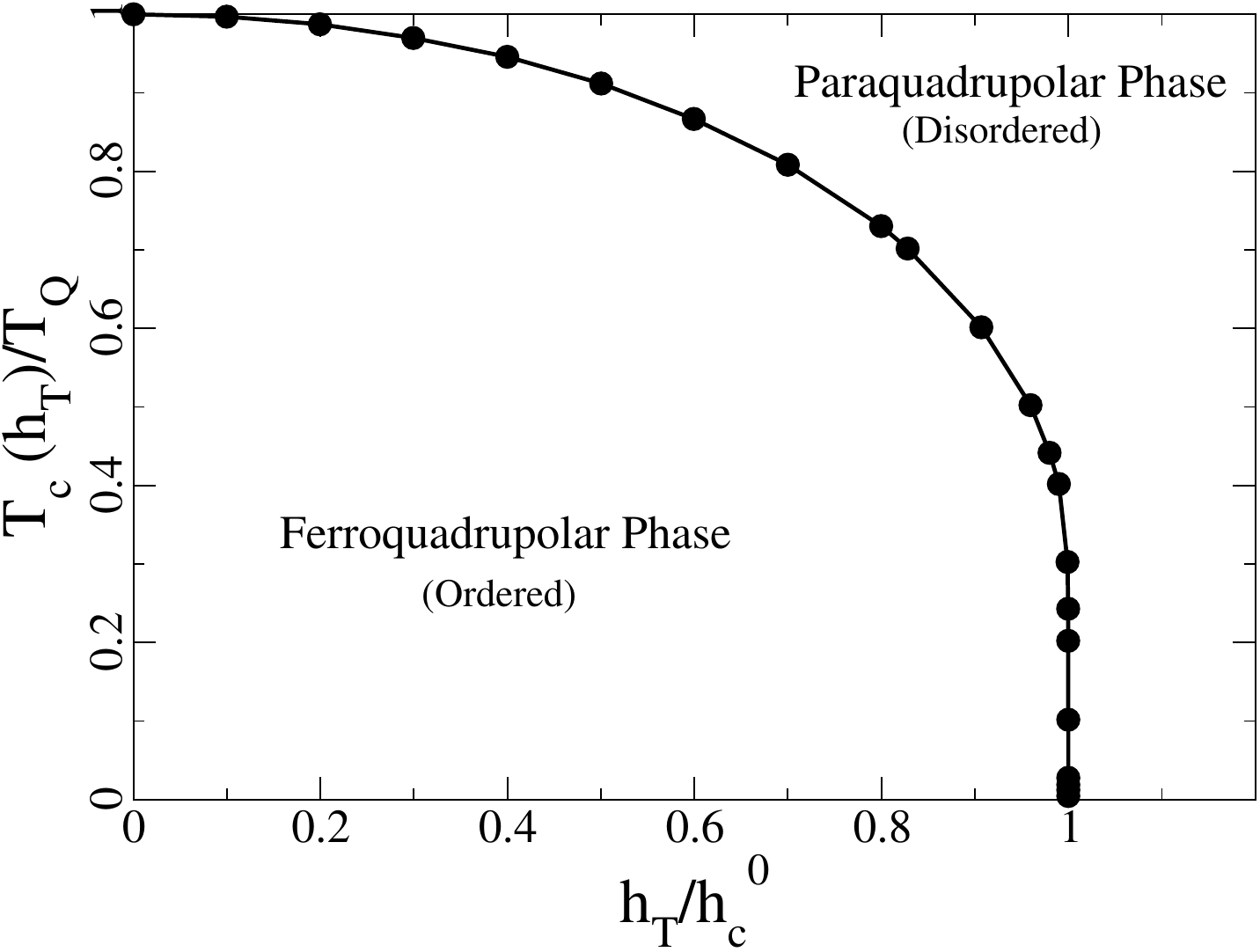}
    \caption{In the absence of the nuclear coupling $A=0$, the phase diagram is determined using Binder cumulant crossings for $N = 32, 64, 128, 256, 512, 1024$. The phase boundary agrees well with mean-field theory for the transverse-field Ising model and with the experimental results of Refs.~\cite{massat2022field,curro2024quantum}.}
    \label{fig:phase_diagram}
\end{figure}

We also calculate the transverse susceptibility, defined as 
\begin{equation}
\chi_{xx} = \pdv{\langle S^x \rangle}{h_x},
\end{equation}
where $S_x$ is the total pseudospin operator along the x-direction.

For clarity, we summarize the numerical approaches used for the results presented below. The exact-diagonalization procedure based on the decomposition into total electronic-pseudospin sectors and, for finite $A$, conserved nuclear-spin sectors is described above in this section. For $A=0$, the thermodynamic and dynamical quantities are calculated using this exact-diagonalization approach, while mean-field theory is used for comparison. For finite $A$, the thermodynamic quantities are obtained by summing over the nuclear-spin sectors described above, with the adaptive nuclear-sector selection scheme of Appendix~\ref{App:D} employed for the low-temperature phase boundary at large system sizes. The excitation gaps and dynamical structure factors are calculated from the corresponding low-energy eigenstates. The experimental entropy data in Fig.~\ref{fig:entropy_match} are obtained independently by integrating the measured specific heat data of Ref.~\cite{massat2022field}.

\section{Analysis and Benchmarking for $A=0$}
\subsection{Phase Diagram}
In the absence of nuclear coupling ($A=0$), the model in Eq. (\ref{eq:H})  reduces to an infinite-range Ising model with transverse magnetic ($h_x$) and strain ($h_y$) fields and the solution is obtained using exact diagonalization. The total transverse field $h_T$  can be written as $h_T = \sqrt{h_x^2 + h_y^2}$. The phase boundary (Fig.~\ref{fig:phase_diagram}) is invariant for combinations yielding the same $h_T$.  We define $h_c^0$ as the critical field
at $T=0$ and $T_Q$ as the critical temperature at $h_T=0$. For $T<T_c(h_T)$, the doublets are in the ordered phase. For $T>T_c(h_T)$, or at high fields $h_{x/y} > h_c(T)$, the system goes into the paraquadrupolar phase; these results are consistent with previous studies. This transition is second order. Because the $S^z$ order parameter couples directly to the $B_{2g}$ strain, the system goes from  a $D_{4h}$ paraquadrupolar phase to a $D_{2h}$ ferroquadrupolar phase.

\subsection{Entropy and Magnetic Susceptibility}
\begin{figure}[h]
    \centering
    \includegraphics[width=1.0\columnwidth]{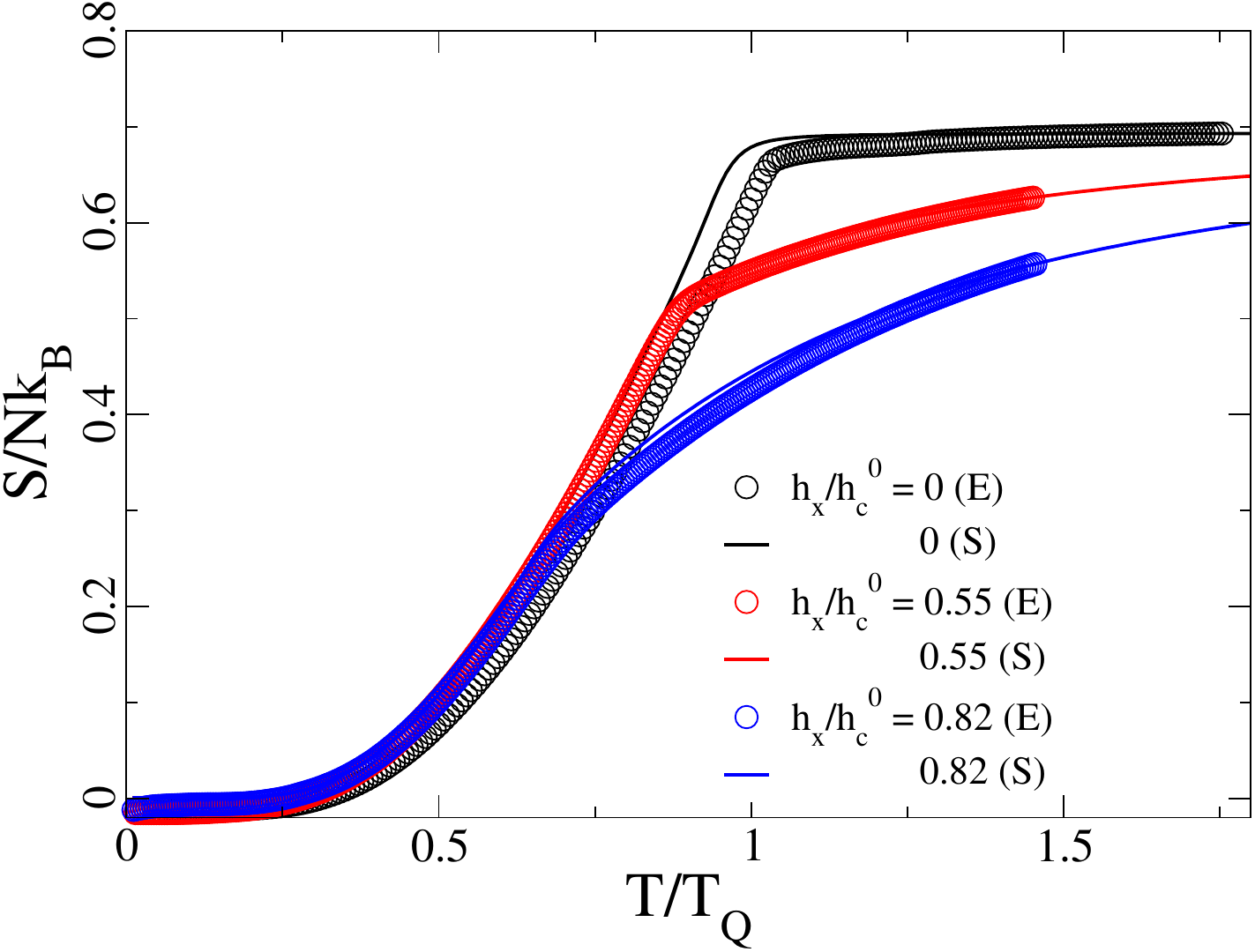}
    \caption{Entropy per site as a function of temperature for three fields $h_x/h_c^0 = 0$, $0.55$, and $0.82$. The solid lines represent the calculated (written as '$S$' in the figure) entropy for $N=1024$, while the circles indicate the experimental data estimated by integrating the experimental specific heat data from Ref. \cite{massat2022field}.}
    \label{fig:entropy_match}
\end{figure}

The fluctuations in the system have both quantum and thermal origins, and the specific heat $C_v$ is proportional to the energy variance $ \braket{E^2}-\braket{E}^2$, whereas the entropy encodes the information about the available states of the system. The specific heat $C_v$ has been measured experimentally as a function of $T$ and transverse magnetic field $h_x$ \cite{massat2022field}. We compute the entropy by integrating the specific heat data, obtained from the experimental measurements of heat capacity in Ref. \cite{massat2022field}, using Simpson’s $3/8$ rule; the resulting values are shown in circles in Fig. \ref{fig:entropy_match}. The solid lines show the ED results for the model. The resulting experimental entropy, shown in Fig. \ref{fig:entropy_match}, is in good agreement with the results of the model.

\begin{figure}[h]
    \centering
    \includegraphics[width=1.0\linewidth]{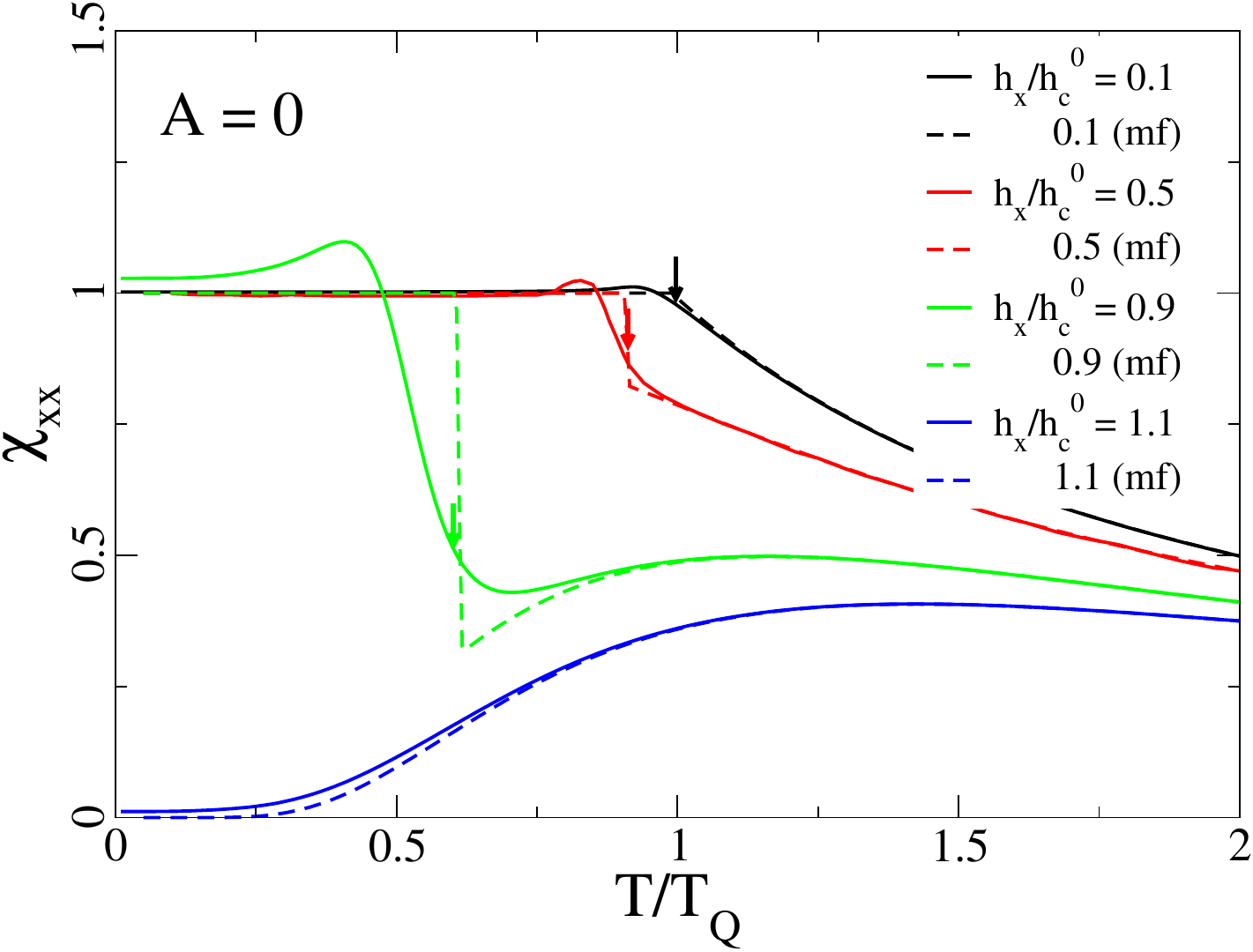}
    \caption{Temperature dependence of the transverse susceptibility $\chi_{xx}$ for $h_x/h_c^0=0.1$, $0.5$, $0.9$, and $1.1$ for $N=1024$. For $h_x/h_c^0=0.9$, $\chi_{xx}$ exhibits a pronounced peak at $T/T_Q\simeq0.48$, close to the ferroquadrupolar transition temperature obtained from the Binder-cumulant analysis. The arrows indicate the corresponding transition temperatures. Dashed lines show the mean-field results.}
    \label{fig:chi_xx}
\end{figure}

To understand the quantum fluctuation effects in more detail, we first study the transverse magnetic susceptibility $\chi_{xx}$ as a function of temperature 
for different transverse fields: in the ordered phase at small $h_x/h_c^0=0.1$, $h_x/h_c^0=0.5$, closer to the phase transition point but still in the ordered phase at $h_x/h_c=0.9$,  and in the disordered phase at $h_x/h_c^0=1.1$  
as shown in Fig.~\ref{fig:chi_xx}. In mean-field theory, $\chi_{xx}$ in the ordered state remains a constant, jumps at the transition, and then gradually  decreases with temperature in the paraquadrupolar phase. In contrast, our results show a weak temperature dependence of the susceptibility in the ordered phase. It initially increases with temperature, passes through a weak maximum, and then decreases rapidly near the transition. Finally, it
gradually decays with temperature in the disordered state. Closer to the transition, at $h_x/h_c^0=0.9$, $\chi_{xx}$ shows a peak and then a sharp decrease at  $T/T_Q=0.48$, followed by a slight upturn and then a gradual decrease. The peak in $\chi_{xx}$ near the transition point is a signature of quantum fluctuations in the system. In the disordered phase, a gap opens up, and therefore, the system requires a finite temperature to access the magnetic states. This leads to vanishing $\chi_{xx}$ at $T=0$ and activated behavior at finite temperatures as shown in Fig. \ref{fig:chi_xx}.


\section{In the presence of nuclear coupling}
\subsection{Phase Diagram and Entropy}

\begin{figure}[h]
    \centering
    \includegraphics[width=1.0\columnwidth]{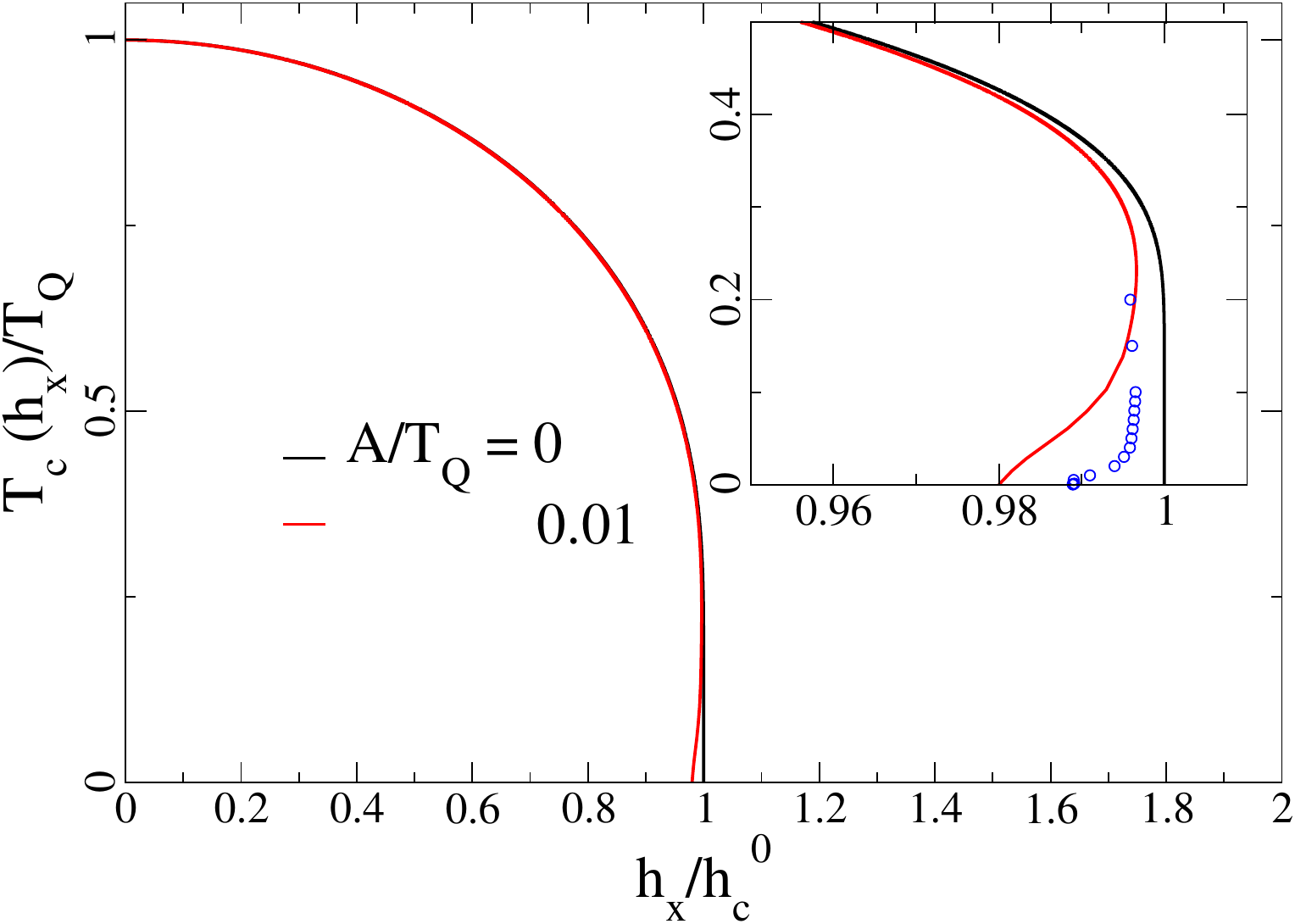}
    \caption{Phase diagram for $A/T_Q=0$ (black) and $A/T_Q=0.01$ (red) obtained from mean-field theory. The inset provides a zoomed-in view near the quantum critical point. Exact diagonalization (blue circles with error bars), obtained from finite-size extrapolation of the minima of $d\langle (S^z)^2\rangle/dh_x$, predicts a smaller bend-back of the phase boundary than mean-field theory. For $A/T_Q=0.01$, the mean-field bend-back is approximately $2\%$, whereas the exact diagonalization result is about $1\%$.}
    \label{fig:boundary_shift_A}
\end{figure}

We now focus on the effects of the hyperfine coupling $A$ on the electronic phase diagram in the $h_x$--$T$ plane (with $h_y=0$). As shown by the mean-field analysis in Appendix~\ref{App:C}, the hyperfine interaction polarizes the nuclear spins, which in turn generates an effective transverse field acting on the electronic pseudospins. This feedback reduces the critical external field at low temperatures. The mean-field and exact diagonalization (ED) results are compared in Fig.~\ref{fig:boundary_shift_A}. Both approaches predict an inward bend-back of the phase boundary relative to the $A=0$ case. However, quantum fluctuations substantially modify the mean-field picture. In addition to reducing the magnitude of the bend-back at $T=0$ from approximately $2\%$ in mean-field theory to about $1\%$ for the relevant $A$ value, ED results show significant qualitative changes for the finite-temperature phase boundary. Whereas mean-field theory predicts an extended low-temperature regime over which the phase boundary bends from the $A=0$ result, quantum fluctuations confine most of the deviation to temperatures much closer to $T=0$. Thus, quantum fluctuations not only reduce the magnitude of the hyperfine-induced shift but also suppress its development over much of the low-temperature phase diagram. The ED phase boundary is obtained by finite-size extrapolation of the minima of $d\langle(S^z)^2\rangle/dh_x$, computed using the adaptive nuclear-sector selection algorithm described in Appendix~\ref{App:D}. The resulting critical field is located at $h_x/h_c^0\simeq0.99$. As shown in the following section, the excitation gap closes at the same field, providing an independent dynamical signature of the quantum phase transition.

\begin{figure}[h]
    \centering
    \includegraphics[width=1.0\columnwidth]{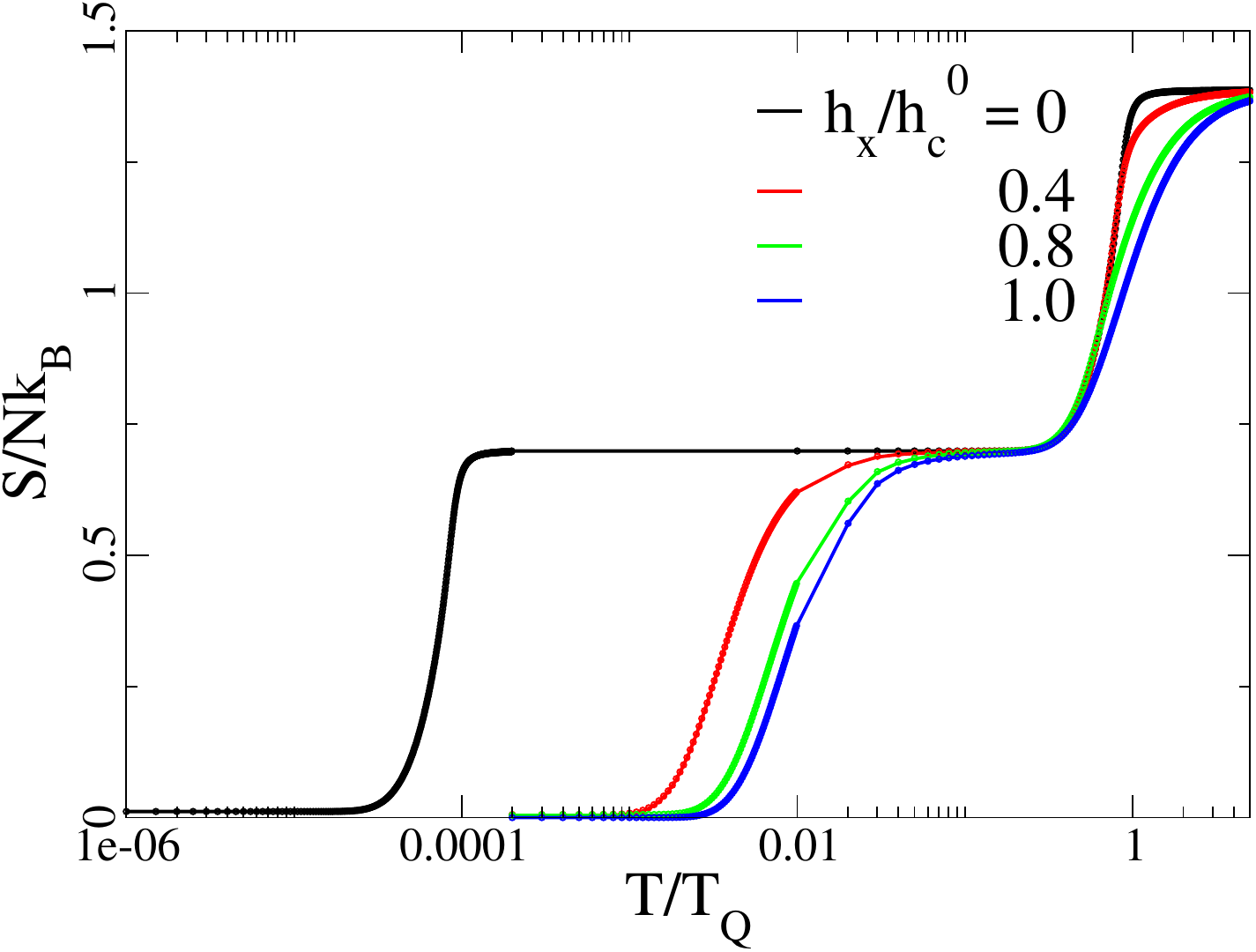}
    \caption{
    Entropy per site $S/Nk_B$ as a function of $T/T_Q$ for
    $h_x/h_c^0=0$, $0.4$, $0.8$, and $1$ at $A/T_Q=0.01$.
    The temperature axis is shown on a logarithmic scale. For $h_x=0$, two entropy-release features are visible, associated with the nuclear-spin ordering at $T_N$ and the electronic transition near $T_Q$. For finite $h_x$, the low-temperature nuclear transition is replaced by a crossover. Solid lines correspond to $N=160$, while symbols correspond to $N=120$; the two system sizes are indistinguishable within the resolution of the plot.
}
    \label{fig:entropy_hx}
\end{figure}

We also explore the entropy per site $S/N$ of this model in the presence of the nuclear coupling using the $ED$ calculations. The results are shown in  Fig. ~\ref{fig:entropy_hx} for four values of $h_x/h_c^0=0$, $0.4, 0.8$, and $1$. For $h_x/h_c^0=0$, $S/N$ shows two sharp transitions; the lower temperature transition is associated with nuclear-spin ordering at a critical temperature $T_N$, whereas the second transition indicates the electronic FQ to PQ phase transition at $T=T_Q$. The entropy per site $S/N$ increases from $0$ to $\ln2$ at the lower temperature transition at $T_N$, where nuclear spins become disordered while the electronic pseudospins are still in the ordered state. The second sharp increase in entropy happens around $T=T_Q$. At this temperature, electronic order vanishes; at higher temperatures $S/N$ approaches $\ln 4$. In this regime, the electronic pseudospin and nuclear-spin degrees of freedom together provide four states per site. For $h_x\neq0$, the transverse field polarizes the electronic pseudospins along $x$, thereby generating an effective field on the nuclear spins through the hyperfine interaction. Consequently,
upon increasing temperature at fixed $h_x$, the nuclear polarization
is gradually lost and the entropy increases continuously from $0$
toward $\ln 2$. This represents a crossover rather than a sharp
nuclear phase transition, whereas the electronic entropy release
remains comparatively narrow around the electronic transition
temperature. At fixed temperature, in contrast, increasing $h_x$
enhances the polarization of the electronic pseudospins and,
through the hyperfine coupling, the nuclear spins, thereby reducing
$S/N$, as seen in Fig.~\ref{fig:entropy_hx}.

\subsection{Energy spectrum}
\begin{figure}[h]
    \centering
    \includegraphics[width=1.0\columnwidth]{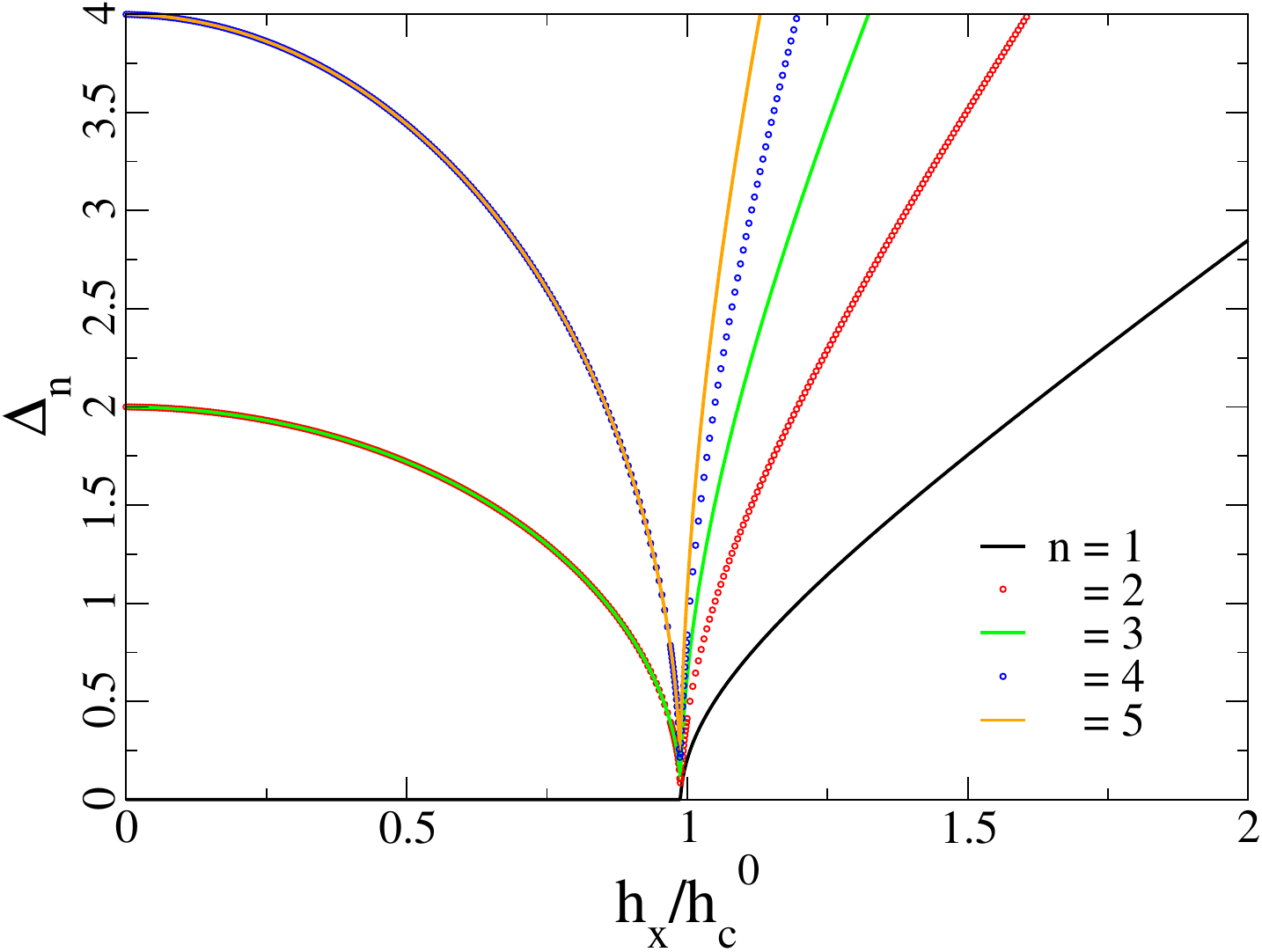}
    \caption{Energy gap $\Delta_n$ as a function of the transverse field $h_x$, corresponding to the energy difference between the ground state and the $n$th excited state for system size $N = 4\times10^5$ and nuclear spin coupling $A = 0.01$. In the ordered phase (left side of the plot), the low-energy excitation branches exhibit near-degeneracies associated with the doubly degenerate ground state. In particular, the $n=2$ and $n=3$ branches, as well as the $n=4$ and $n=5$ branches, overlap within the plotted resolution. The lowest excitation gap closes near the quantum critical point at $h_x/h_c^0 \simeq 0.99$ and reopens in the disordered phase, while the higher excitation gaps increase more rapidly with $h_x$.}
    \label{fig:energy_gap}
\end{figure}

We next analyze the energy spectrum, which governs the low-temperature thermodynamic properties. Fig. \ref{fig:energy_gap} shows the five lowest excitation gaps. In the ordered phase and for $h_x <h_c^0$, the shown spectrum is approximately twofold degenerate. To examine the structure of the low-lying excitation spectrum in the ordered phase, we set $h_x = 0$, $h_y=0$, and $A=0$ in the Hamiltonian in Eq. (\ref{eq:H}), which can be expressed in terms of the total spin as
\begin{equation}
H = -\frac{2J_z}{N} \left( S_{\mathrm{tot}}^{z} \right)^2 .
\end{equation}

In this limit, the ground state is fully polarized with $S_{\mathrm{tot}}^{z} = \pm N/2$, leading to a twofold degenerate ground state. 
A single spin-flip excitation reduces the total magnetization to $S_{\mathrm{tot}}^{z} = N/2 - 1$, and the low-lying  excitation gap $\Delta_n$, where $n$ labels $n$th excited state, can be derived from the energy difference
\begin{equation}
\Delta_n = \frac{2J_z}{N} \left[ \left(\frac{N}{2}\right)^2 - \left(\frac{N}{2}-n\right)^2 \right] 
= 2nJ_z\left(1 - \frac{n}{N}\right).
\end{equation}

\begin{figure*}[t]
    \centering

    \begin{minipage}{0.48\linewidth}
        \centering
        \includegraphics[width=\linewidth]{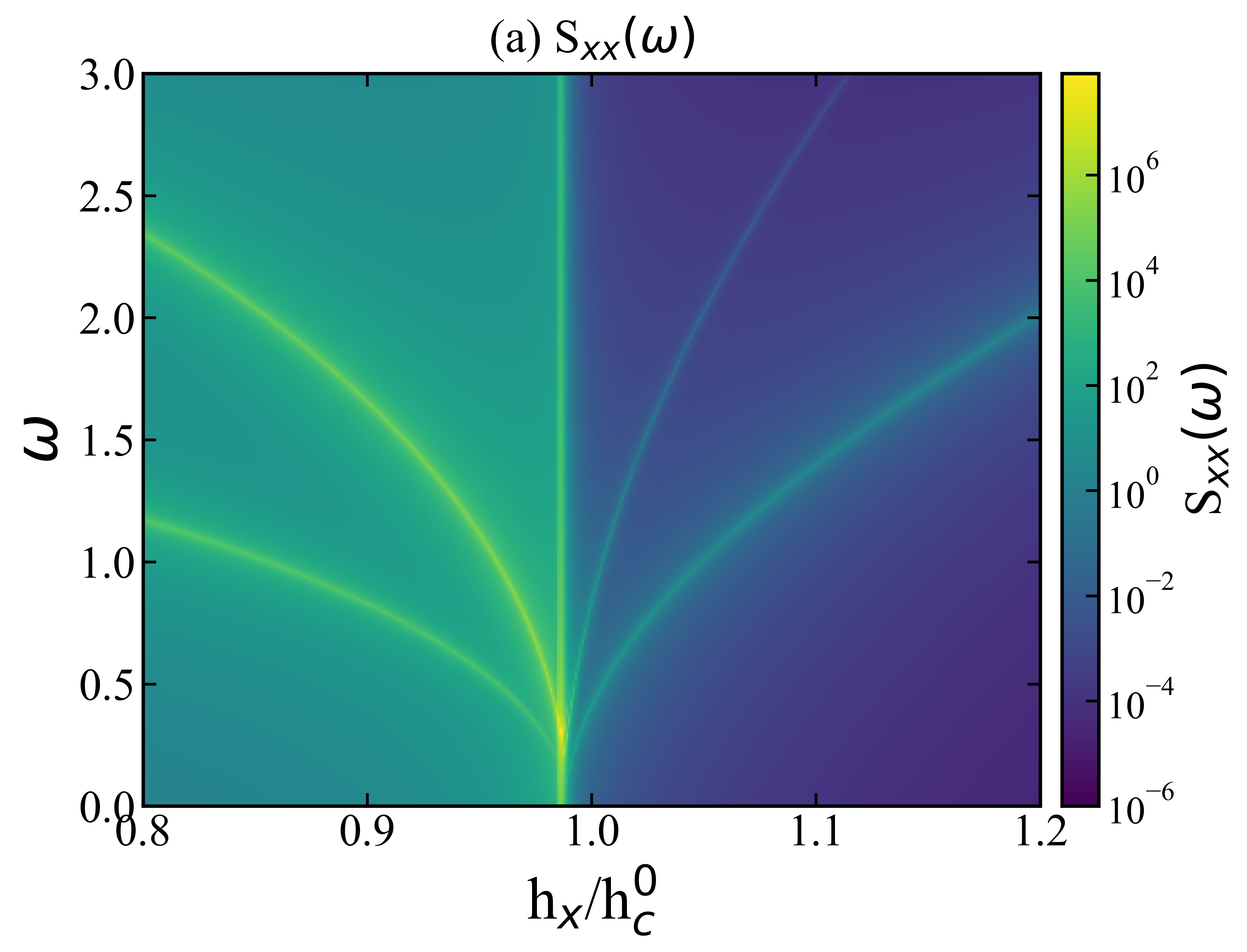}
    \end{minipage}
    \hfill
    \begin{minipage}{0.48\linewidth}
        \centering
        \includegraphics[width=\linewidth]{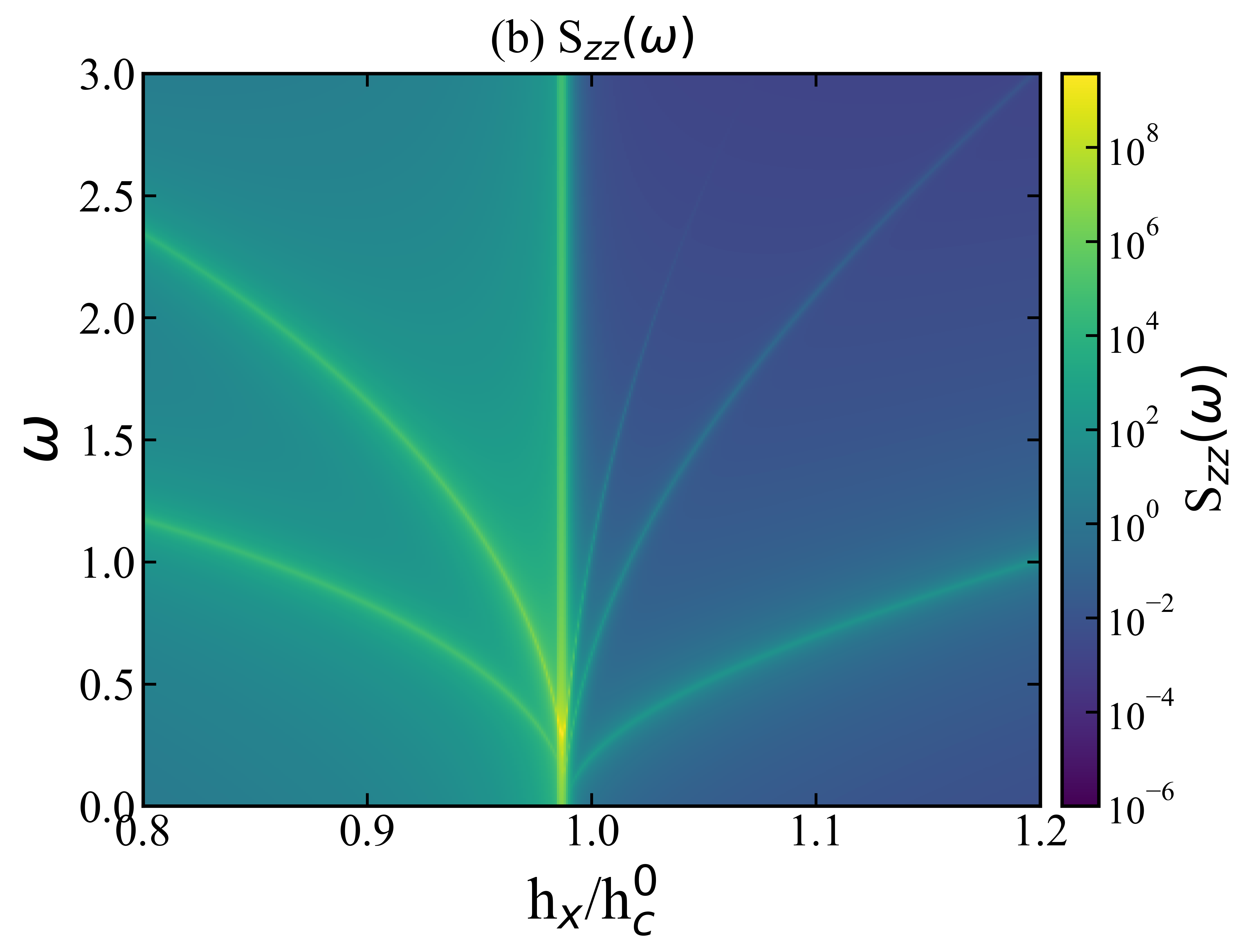}
    \end{minipage}

    \vspace{0.4cm}

    \begin{minipage}{0.48\linewidth}
        \centering
        \includegraphics[width=\linewidth]{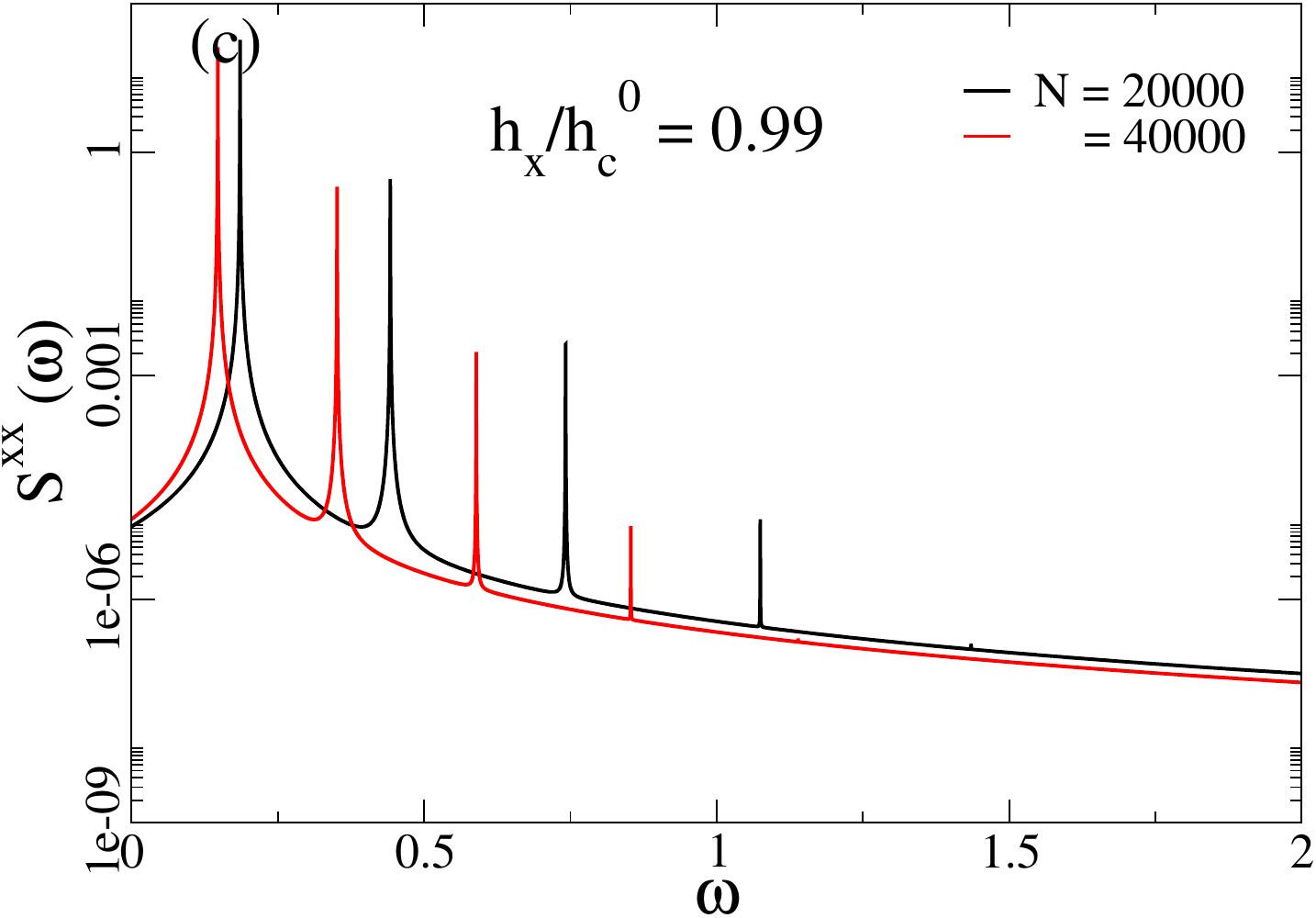}
    \end{minipage}
    \hfill
    \begin{minipage}{0.48\linewidth}
        \centering
        \includegraphics[width=\linewidth]{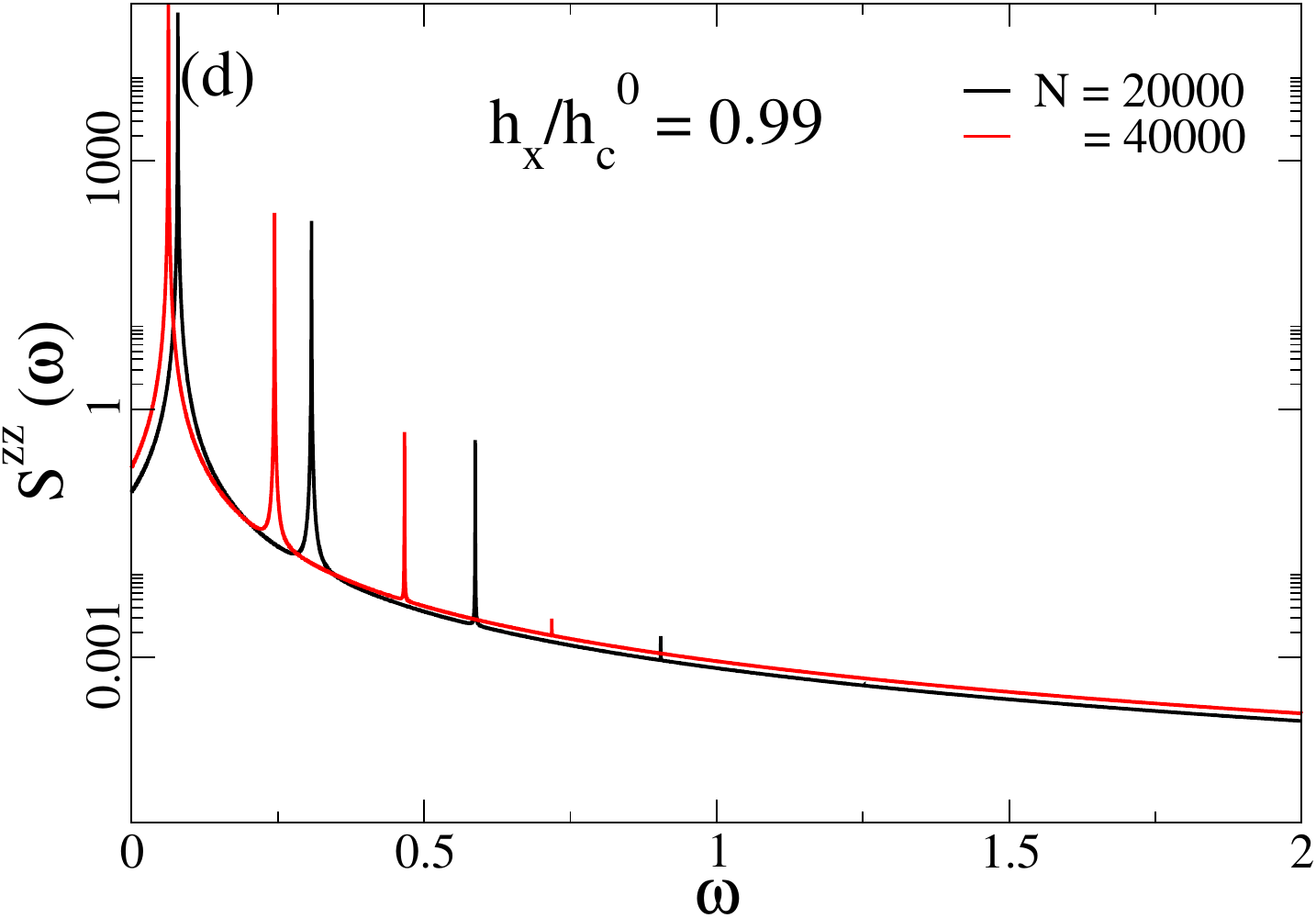}
    \end{minipage}

    \caption{Dynamical structure factors per site for $N=4\times10^5$ and $A/T_Q=0.01$ with $\eta=10^{-3}$. Here, $(1\!-\!n)$ denotes the transition from the ground state to the $n-1$th excited state, with the corresponding spectral weight determined by the matrix element $|\langle\psi_n|S^\alpha|\psi_{\rm gs}\rangle|^2$. To make the weak spectral branches visible in panels (a) and (b), selected matrix elements are rescaled for visualization: $(1\!-\!2)$ is unchanged, $(1\!-\!3)$ is multiplied by $10^3$, $(1\!-\!4)$ by $10^5$, and $(1\!-\!5)$ by $10^7$. These rescaling factors are used only for visualization and do not represent the physical relative spectral weights. (a) Color map of $S^{xx}(\omega)$ as a function of frequency $\omega$ and transverse field $0.8<h_x/h_c^0<1.2$. (b) Same as panel (a), but for $S^{zz}(\omega)$ for $0.8<h_x/h_c^0<1.2$. (c) $S^{xx}(\omega)$ as a function of $\omega$ at fixed $h_x/h_c^0=0.99$ for $\omega\leq2$. (d) $S^{zz}(\omega)$ as a function of $\omega$ at the same fixed $h_x/h_c^0$ for $\omega\leq2$.}

    \label{fig:spectral_function}
\end{figure*}

For $N \rightarrow \infty $  and for any finite $n$, the single-spin-flip excitation gap is $\Delta= 2J_z$,  whereas two spin-flip excitations cost $\Delta_2=4J_z$. These excitations form degenerate multiplets in the ordered phase, consistent with the observed degeneracy of the lowest excitation gaps. Upon increasing $h_x$, the transverse field mixes different magnetization sectors, progressively reducing the excitation energies and lifting the degeneracies. These gaps decrease as $h_x$ increases and vanish at the critical point. In the PQ phase, the ground state is non-degenerate and higher excitations no longer form doublets; five distinct gaps are shown in Fig. \ref{fig:energy_gap}.

\subsection{Signatures of the Excitation Spectrum}
The gaps alluded to above can be probed using inelastic neutron scattering by measuring the dynamical structure factor (DSF)  $S^{\alpha\alpha}(q=0,\omega)$ defined in Eq. (\ref{eq:spectral_func}). We compute transverse $S^{xx}(q=0,\omega, h_x)$ and longitudinal $S^{zz}(q=0,\omega, h_x)$ structure factors, shown in Fig. \ref{fig:spectral_function}. 
In the ordered and disordered phases, the spectral weights associated with several higher excited modes become extremely small compared to the dominant low-energy branch. As a result, without rescaling, only the strongest excitation branch would remain visible in the dynamical structure factor plots. To visualize the evolution of multiple excitation branches across the phase diagram and facilitate comparison with the corresponding gap structure shown in Fig. \ref{fig:energy_gap}, selected weak spectral features have been multiplied by the factors indicated in the caption. We emphasize that these rescaled branches correspond to much smaller physical intensities than the dominant mode. The variations of $S^{xx}$ and $S^{zz}$ with $h_x/h_c^0$ look similar but the intensities are different. In  Figs.~\ref{fig:spectral_function}(a) and ~\ref{fig:spectral_function}(b), the spectral functions are shown across the phase diagram, and the quantum critical point is clearly identified near $h_x/h_c^0 \simeq 1$ at which the gap vanishes. In the absence of nuclear coupling the transition point shifts towards higher fields, and $\Delta_1$ also shifts accordingly. In Figs.~\ref{fig:spectral_function}(c) and ~\ref{fig:spectral_function}(d), we show the frequency dependence of $S^{xx}(\omega)$ and $S^{zz}(\omega)$ for $N=2\times10^5$ and $N=4\times10^5$ at $h_x/h_c^0=0.99$. Both structure factors exhibit multiple low-frequency peaks. The number of states carrying finite spectral weight increases with system size, although many of the additional states possess very small weights and are not readily visible in the spectra. This behavior is consistent with an increasing density of low-energy excitations in the thermodynamic limit.

The absence of clearly visible spectral weight at higher frequencies does not imply the absence of higher excited states; rather, the matrix elements of the collective $q=0$ spin operators for these states are extremely small. The weak high-frequency tails visible on the logarithmic scale are primarily generated by the Lorentzian broadening $\eta=10^{-3}$ used to represent the discrete spectral lines.

The behavior of the DSF is consistent with the energy-gap analysis shown in Fig.~\ref{fig:energy_gap}. In the ordered phase, the ground state is doubly degenerate, whereas this degeneracy is lifted in the disordered phase and a finite gap opens and increases deeper in the disordered phase. At the critical field $h_x/h_c^0=0.99$, the energy gap between the
ground state and the first excited state exhibits an inverse
power-law scaling with system size, $\Delta \sim N^{-\alpha}$, with $\alpha \simeq 1/3$, as detailed in Appendix~\ref{App:E}. The vanishing of the gap in the thermodynamic limit is consistent with the expected behavior at a continuous quantum critical point. The observed $N^{-1/3}$ scaling therefore serves as a quantitative consistency check of the finite-size critical behavior of the infinite-range transverse-field Ising model.

\begin{figure}[h]
    \centering
    \includegraphics[width=1.0\linewidth]{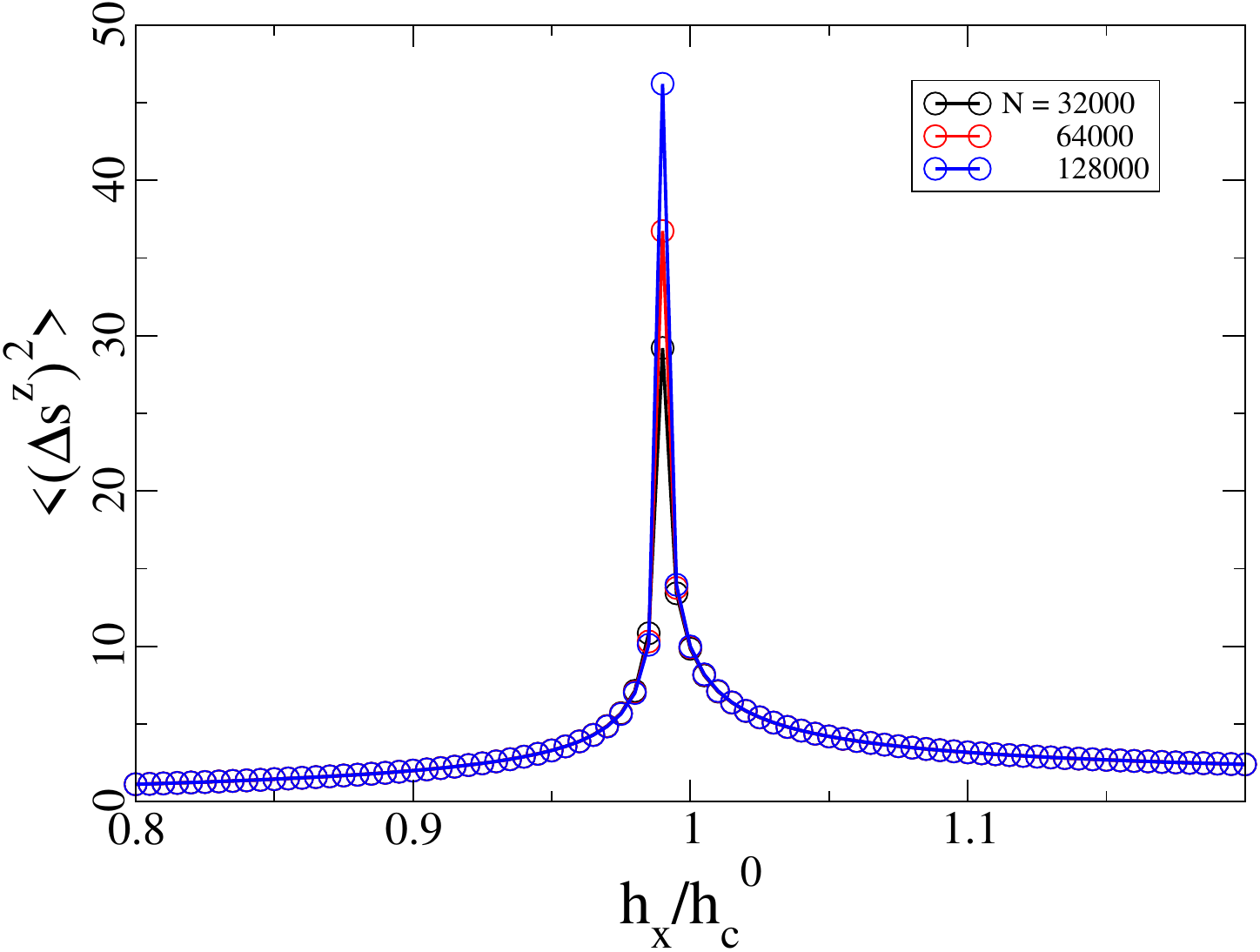}
    \caption{Transverse-field dependence of the longitudinal spin fluctuation 
    $\langle (\Delta s^{z})^{2} \rangle$ for different system sizes, $N=32000$, $64000$ and $128000$. 
    A pronounced peak appears at the critical field $h_x/h_c^0 \approx 0.99$, 
    indicating enhanced quantum fluctuations in the vicinity of the 
    quantum phase transition. The sharp maximum reflects the critical 
    enhancement of longitudinal spin variance as the system approaches 
    the critical transverse field from both sides. The peak heights increase with system size, suggesting a divergence in the thermodynamic limit. The transition point appears to be nearly system-size independent.}
    \label{fig:gs_qfi}
\end{figure}

The system is intrinsically many-body, and its zero-temperature quantum fluctuations can be quantified through the variance of the spin operator $\langle \Delta O^2 \rangle$ in the ground state. The variance of an operator $\hat{O}$ can be defined as  
\begin{equation}
\langle \Delta O^2 \rangle 
= 4\left( 
\langle \psi | \hat{O}^2 | \psi \rangle 
- \langle \psi | \hat{O} | \psi \rangle^2 
\right).
\end{equation}
The ground state belongs to the sector characterized by $m=-N/2$ and $s_t=N/2$, and the corresponding ground-state wave function $\ket{\psi}$ is obtained from this sector.

We calculate the variance of the longitudinal component of spin $\langle (\Delta s^z)^2 \rangle$ as a function of $h_x$ (shown in Fig. \ref{fig:gs_qfi}), and the variance increases with field, reaches a maximum at the transition, and decreases thereafter. 
We present results for three system sizes, $N=32000$, $64000$, and $128000$, and observe that the peak consistently occurs at the critical point $h_x/h_c^0 = 0.99$. The critical field is essentially independent of system size. In contrast, the peak height increases with system size, indicating a divergence of $\langle (\Delta s^z)^2 \rangle$ in the thermodynamic limit. 

\subsection{Effect of strain} 
As pointed out earlier, strain provides a means of tuning the FQ phase through its coupling to the crystal-field degrees of freedom. We examine the entropy and nuclear ordering as a function of the strain field $h_y$, setting $h_x=0$ and retaining a finite hyperfine coupling $A \ne 0$. For $h_x=0$, $\langle S^x \rangle $ is zero, but the nuclear spins remain coupled to the electronic degrees of freedom via the hyperfine coupling. Via second-order perturbation theory, this generates an effective coupling between the nuclear spins, causing the nuclear spins to order at the low temperature $T_N$. It is important to note that the effects of the strain field $h_y$ and the transverse magnetic field $h_x$ on the nuclear degrees of freedom are qualitatively different. The origin of this difference lies in the form of the hyperfine coupling, which couples the nuclear spins specifically to the electronic $S^x$ component. A finite transverse magnetic field $h_x$ directly induces a nonzero $\langle S^x \rangle$, generating an effective field acting on the nuclear spins through the hyperfine interaction. As a result, the nuclear ordering transition is replaced by a smooth crossover. In contrast, the strain field $h_y$ couples to the orthogonal $S^y$ component and does not directly induce a finite $\langle S^x \rangle$. Consequently, the nuclear spins interact only through virtual electronic fluctuations, leading to an effective interaction of order $A^2/h_y$ (in the $h_y >> J_z$ limit) and allowing a genuine low-temperature nuclear ordering transition to survive.

\begin{figure}[h]
    \centering
    \includegraphics[width=1.0\columnwidth]{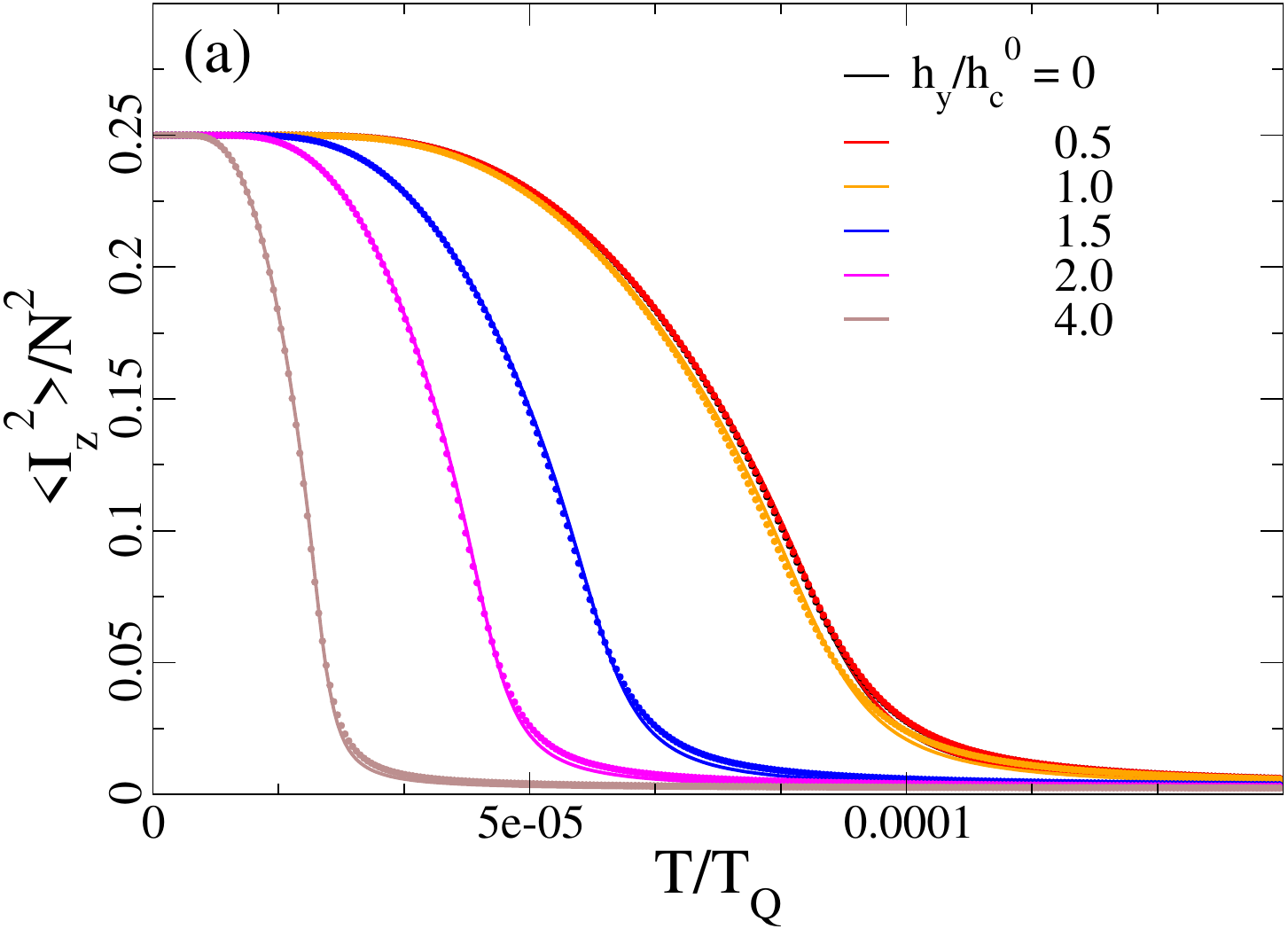}
    \includegraphics[width=1.0\columnwidth]{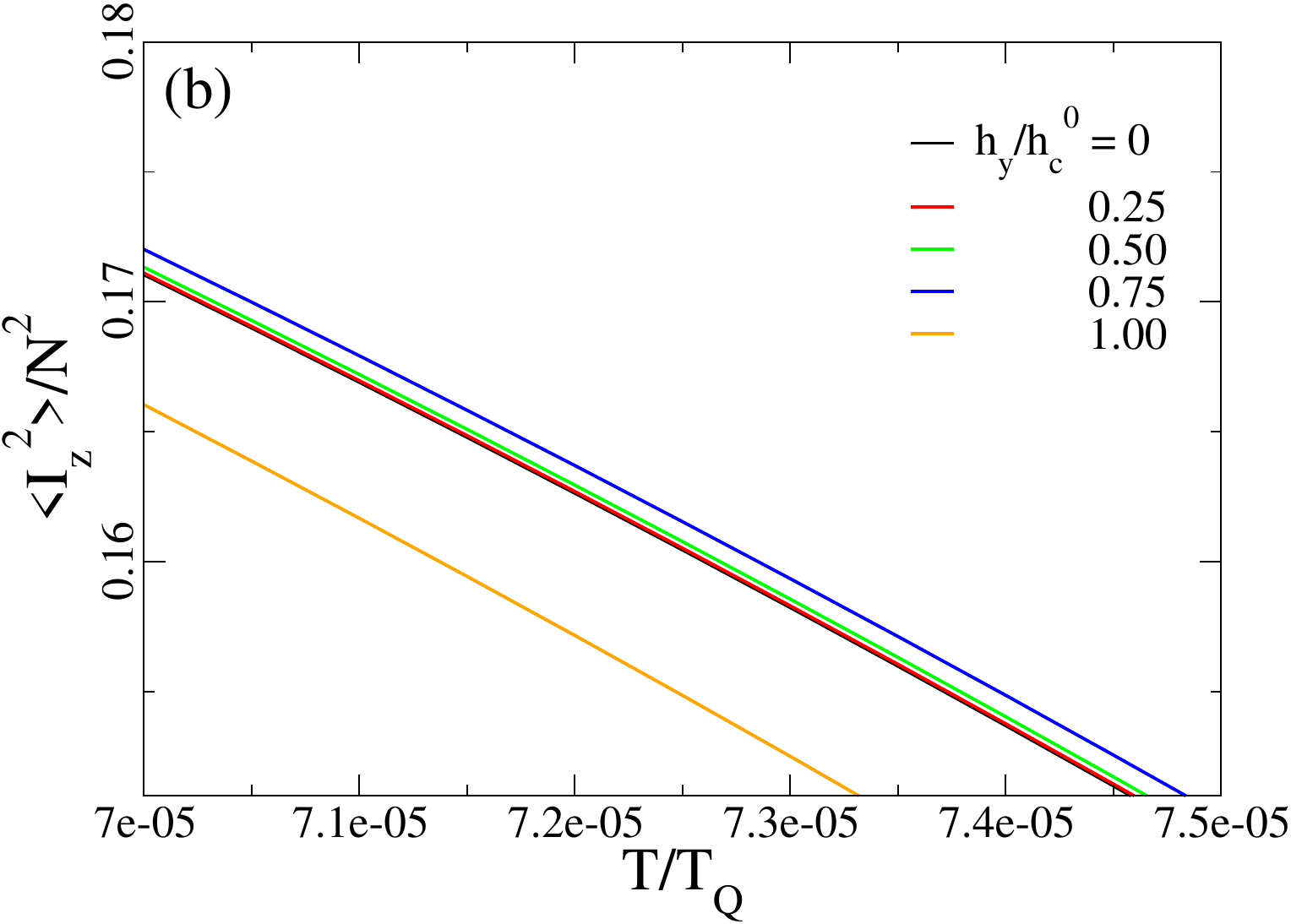}
    \caption{Temperature dependence of the normalized nuclear expectation
$\langle I_z^2\rangle/N^2$ for various values of $h_y/h_c^0$ at
$A/T_Q=0.01$. (a) Full temperature dependence over the complete
strain-field range. The nuclear ordering scale varies only weakly
with $h_y$ at small strain. Solid lines correspond to $N=160$ and circles to $N=120$. (b) Expanded view of the low-temperature
region for intermediate strain fields, resolving the nonmonotonic
dependence of the characteristic nuclear ordering temperature on
$h_y$.}
    \label{fig:Iz2_hy_A}
\end{figure}

In Fig. \ref{fig:Iz2_hy_A}(a), the effect of the strain field $h_y$ on the normalized nuclear expectation $\langle I_z^2\rangle/N^2$ is shown as a function of $T/T_Q$. The characteristic nuclear ordering temperature exhibits a nonmonotonic dependence on strain: it initially increases with $h_y$, reaches a maximum, and subsequently decreases for larger $h_y$. The weak variation at small $h_y$ is more clearly resolved in the expanded intermediate-field range shown in Fig.~\ref{fig:Iz2_hy_A}(b).
 

\begin{figure*}[t]
    \centering
    \includegraphics[width=1.0\textwidth]{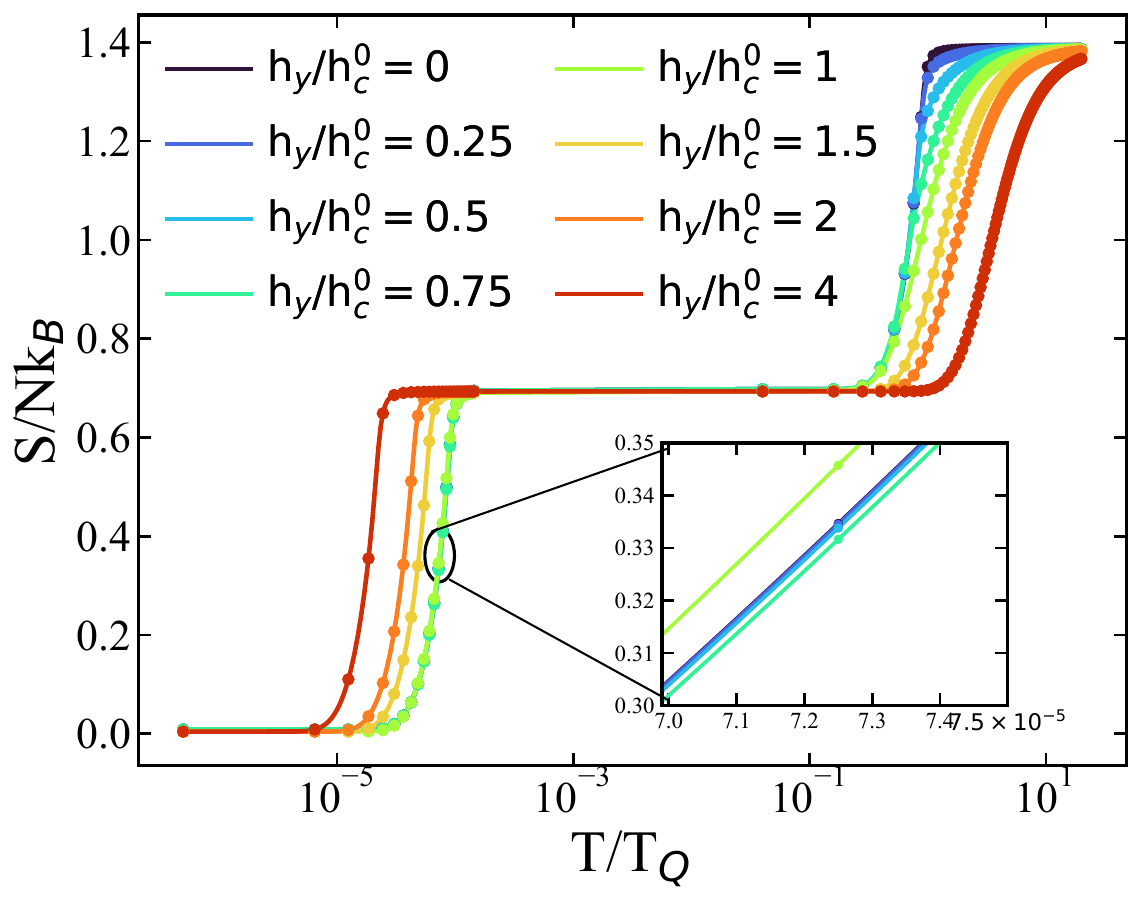}
    \caption{Entropy per site $S/Nk_B$ as a function of $T/T_Q$ showing two transitions at $T_c (h_y)$ and $T_N$. The nuclear transition depends nonmonotonically on the strain field $h_y$. Higher $h_y/h_c^0$ shifts $T_N$ to much lower values. (The parameters are $A/T_Q = 0.01$, $N=160$. Circles are used for $N=120$.) \textbf{Inset}: Zoom of the low-temperature entropy-release region for selected strain fields. Since the curves nearly overlap, their colors are explicitly identified as black ($h_y/h_c^0 = 0$), royal blue ($0.25$), sky blue ($0.5$), turquoise ($0.75$), and lime green ($1$).}
    \label{fig:entropy_hy}
\end{figure*}

We analyze the entropy as a function of $T/T_Q$ for different values of $h_y$. The entropy $S/N$ (Fig.~\ref{fig:entropy_hy}) shows two sharp jumps: at $T_N$ ($0 \to \ln2$) and $T_c (h_y)$ ($\ln2 \to \ln4$). Similar to previous results, the first jump represents the nuclear spin transition, whereas the second jump corresponds to the electronic transition. The electronic doublets and nuclear spins couple through virtual electronic fluctuations. 
The strain dependence of the entropy also indicates the possibility of an elastocaloric response under adiabatic strain sweeps. In particular, at fixed entropy near the electronic entropy-release region, increasing $h_y$ shifts the isentropes to higher temperatures, whereas decreasing $h_y$ results in cooling. In contrast, near the nuclear entropy-release region, the entropy curves exhibit a nonmonotonic dependence on strain, as shown in the inset of Fig.~\ref{fig:entropy_hy}. For clarity, the inset identifies the nearly overlapping curves by color: black ($h_y/h_c^0=0$), royal blue ($0.25$), sky blue ($0.5$), turquoise ($0.75$), and lime green ($1$). In this nonmonotonic regime, increasing the strain field can also produce a heating effect. Beyond this point, further increasing the strain field results in a cooling effect. Although a detailed analysis of the elastocaloric coefficient is beyond the scope of the present work, Fig. \ref{fig:entropy_hy} indicates that strain tuning may provide a route for entropy manipulation at low temperatures.  

\section{Conclusion} 
In this work, we study an effective infinite-range Ising model for TmVO$_4$\cite{massat2022field}. The model describes interactions among the low-energy electronic pseudospins associated with the non-Kramers doublets and
their hyperfine coupling to the nuclear spins. We use mean-field theory and exact diagonalization for systems with thousands of spins to calculate thermodynamic and dynamical observables. The phase boundaries are determined partly using the Binder-cumulant analysis and we construct the phase diagram in parameter space of transverse magnetic ($h_x$) and $B_{1g}$ shear‐strain ($h_y$) fields. We determine the effects of the hyperfine coupling $A$ on the phase diagram. The low-temperature phase boundary for the latter case is obtained using extrapolations of the peaks in the $d\langle (S_z)^2 \rangle/dh_x$ data for large system sizes. 

In the pure electronic limit ($A=0$), we recover the ferroquadrupolar-paraquadrupolar transition at $T=T_c(h_x,h_y)$ using Binder-cumulant crossings, which is also marked by a drop in entropy per site from $\ln4$ to $\ln2$, consistent with mean-field theory \cite{massat2022field}. However, quantum and thermal fluctuations near the phase transition modify this behavior.   Introducing a weak hyperfine interaction ($A/T_Q=0.01$) shifts the phase boundary inward at $T=0$ because the nuclear spins polarize, inducing an additional effective transverse field. In mean-field theory, the resulting shift is $\Delta h_x = -2A \tanh \left( \frac{A}{T} \tanh \left( \frac{h_x}{T} \right) \right)$, corresponding to a $2\%$ shift at $T=0$ for $A/T_Q=0.01$. Our exact diagonalization results show that mean-field theory overestimates this bend-back by a factor of approximately $2$. We further demonstrate that quantum criticality is reflected in enhanced transverse susceptibility, a pronounced peak in longitudinal spin variance, and the softening of low-energy excitation gaps. The dynamical structure factor shows a clear closing and reopening of the excitation gap across the quantum critical point, providing experimentally testable signatures for spectroscopic probes.

In the presence of a transverse field ($h_x$), the system does not exhibit a sharp nuclear transition; instead, nuclear moments polarize smoothly, and the residual entropy $\ln2$ is released via a gradual crossover.  By contrast, when a $B_{1g}$ shear strain field ($h_y$) is applied in the absence of $h_x$, we identify a genuine nuclear phase transition at which the remaining entropy collapses to zero and above which the nuclear expectation $\langle I_z^2\rangle/N^2$ vanishes with increasing $N$.  Moreover, \(T_N\) shows a nonmonotonic dependence on strain, initially increasing with \(h_y\), reaching a maximum, and then decreasing at larger strain, showing the interplay between elastic tuning and hyperfine-mediated interactions.

Finally, the two well separated entropy-release scales suggest a two-stage elastocaloric response under adiabatic strain sweeps: an entropy release of $\ln 2$ at the electronic transition temperature $T_c(h_y)$, followed by a second entropy release of $\ln 2$ at the nuclear ordering temperature $T_N$. Near the electronic transition, decreasing the strain field produces cooling, whereas in the nuclear entropy-release regime the nonmonotonic strain dependence of the entropy implies that increasing the strain field first results in heating, followed by cooling beyond $h_y/h_c^0=1$. Given the large and unrealistic strain-fields needed for significant elastocaloric response at the nuclear transition, this may not be a practical material to achieve cooling at very low temperatures.

In summary, the present study provides a minimal yet quantitatively useful framework for understanding two-stage ordering, entropy change, and quantum fluctuations in long-range ferroquadrupolar systems with hyperfine coupling. It bridges electronic-pseudospin criticality and nuclear spin physics within a single microscopic model and offers clear experimental predictions for thermodynamic and dynamical probes. 
These results can also be utilized to understand other magnetic systems described by long-range Ising models. 

\section{Acknowledgments}
We thank Nick Curro, Ian Fisher, and Marc Zic for discussions on the project. SG acknowledges financial support from DST-Inspire. We acknowledge the National Supercomputing Mission (NSM) for providing computational resources on PARAM RUDRA at S.N. Bose National Centre for Basic Sciences, implemented by C-DAC and supported by the Ministry of Electronics and Information Technology (MeitY) and Department of Science and Technology (DST), Government of India. AM acknowledges support from the National Centre for Theoretical Sciences in Taiwan. We acknowledge support from the US National Science Foundation (NSF) Grant No. 2201516 under the Accelnet program of the Office of International Science and Engineering (OISE).

\bibliography{ref}
\bibliographystyle{apsrev4-2}

\appendix
\section{\label{App:B}Origin and mathematical expression for combinatorial factors}
The infinite-range model possesses a large number of symmetries. Since the Hamiltonian is expressed entirely in terms of the collective electronic-pseudospin operators \(S_t^x\), \(S_t^y\), and \(S_t^z\), it commutes with the total spin Casimir operator $S_t^2$. As a result, the Hamiltonian becomes block diagonal in relatively small Hilbert-space sectors of dimension $2s_t+1$, which scales linearly with $N$ for the largest-spin sector $s_t=N/2$. The energy levels and properties within each sector depend on the total spin \( s_t \), which can take values from \( 0 \) to \( N/2 \) (assuming \( N \) is even). The Hilbert space dimension in each \( s_t \) sector is \( 2s_t + 1 \). The number of copies of the spin \( s_t \) sector is the number of ways \( N \) spin-\(\frac{1}{2}\) objects can be combined into a total spin \( s_t \). This multiplicity \( d(s_t) \) is one for \( s_t = N/2 \), and for \( s_t < N/2 \), it can be expressed in terms of combinatorial factors 
as:  
\begin{equation}
    d(s_t) = \binom{N}{\frac{N}{2} + s_t} - \binom{N}{\frac{N}{2} + s_t + 1}.
\end{equation}  
Therefore, instead of dealing with an exponentially large Hilbert space of size \( 2^N \), we only need to diagonalize matrices of size \( 2s_t + 1 \) within each total spin sector, which is much smaller. The full Hilbert space dimension can be recovered from the relation:
\begin{equation}
    2^N = \sum_{S_t = 0}^{N/2} (2s_t + 1) d(s_t).
\end{equation}

 Let us now turn to the nuclear coupling. For our Hamiltonian the $z$-component of total nuclear spin is a conserved quantity. This means that the matrix sizes that need to be diagonalized remain unchanged. However, the calculation has to be repeated for different values $m$ of the $z$-component of the total nuclear spin which ranges from \( -N/2 \) to \( N/2 \). The number of ways in which a particular $m$ arises is given by the combinatorial factor:
\begin{equation}
    d(m) = \binom{N}{\frac{N}{2}+m}.
\end{equation}
The combined total Hilbert space dimension of the electronic and nuclear spins can be recovered from the relation:
\begin{equation}
    4^N = \sum_{m=-N/2}^{N/2}d(m)\sum_{s_t = 0}^{N/2} d(s_t)(2s_t + 1).
\end{equation}

\section{\label{App:C}Mean-field analysis at finite transverse field}

We consider the effect of the nuclear coupling on the shift of the phase boundary to leading order in $A$. The starting Hamiltonian is given by
\begin{equation}
\mathcal{H} = \mathcal{H}_0
+4A\sum_i I_i^z\frac{S_t^x}{N},
\end{equation}
where $A$ is the nuclear coupling constant, $S_t^x=\sum_i S_i^x$ is the total electronic-pseudospin in the $x$-direction, and $I_i^z$ is the nuclear spin operator at site $i$.

We rewrite the Hamiltonian as
\begin{equation}
\mathcal{H}
=
\mathcal{H}_0-h_{\rm nuc}\sum_i I_i^z,
\end{equation}
where, within the mean-field approximation,
\begin{equation}
h_{\rm nuc}=-4A\langle S_i^x\rangle.
\end{equation}
With the convention in which the transverse-field term in $\mathcal{H}_0$ is $+2h_xS_t^x$, the electronic pseudospins polarize opposite to the applied field. Since we require the transverse polarization only to determine the leading-order shift of the \(A=0\) phase boundary, we evaluate \(\langle S_i^x\rangle\) on the unperturbed critical line, where the longitudinal order parameter vanishes. Thus,
\begin{equation}
\langle S_i^x\rangle
=
-\frac{1}{2}
\tanh\left(
\frac{T_Q}{T}\frac{h_x}{h_c^0}
\right),
\end{equation}
where $T_Q=T_c(h_T=0)$ and $h_c^0$ is the critical field at $T=0$ in the absence of nuclear coupling, as discussed in the main text. Consequently, the effective nuclear field is
\begin{equation}
h_{\rm nuc}(T,h_x)
=
2A\tanh\left(
\frac{T_Q}{T}\frac{h_x}{h_c^0}
\right).
\end{equation}

This nuclear field produces a nuclear polarization given by
\begin{equation}
\langle I_i^z\rangle
=
\frac{1}{2}
\tanh\left(\frac{\beta h_{\rm nuc}}{2}\right)
=
\frac{1}{2}
\tanh\left[
\beta A
\tanh\left(
\frac{T_Q}{T}\frac{h_x}{h_c^0}
\right)
\right].
\end{equation}

Substituting this polarization back into the original Hamiltonian gives
\begin{equation}
\mathcal{H}
=
\mathcal{H}_0
+
2A
\tanh\left[
\beta A
\tanh\left(
\frac{T_Q}{T}\frac{h_x}{h_c^0}
\right)
\right]S_t^x.
\end{equation}
Thus, the electronic pseudospins effectively experience a modified transverse field given by
\begin{equation}
h_x
+
2A
\tanh\left[
\beta A
\tanh\left(
\frac{T_Q}{T}\frac{h_x}{h_c^0}
\right)
\right].
\end{equation}
This additional effective field implies that a smaller applied field is required to reach the critical point. The phase boundary is therefore shifted from its $A=0$ value by
\begin{equation}
\Delta h_x
=
-2A
\tanh\left[
\beta A
\tanh\left(
\frac{T_Q}{T}\frac{h_x}{h_c^0}
\right)
\right].
\end{equation}

If we numerically evaluate the phase boundary using the $A=0$ relation,
\begin{equation}
T_c(h_x)
=
T_Q
\frac{h_x/h_c^0}
{\tanh^{-1}(h_x/J_z)},
\end{equation}
we obtain a set of points $(h_x,T)$ along the phase boundary. The corresponding shifted phase boundary is then given by
\begin{equation}
(h_x+\Delta h_x,T).
\end{equation}




\section{\label{App:D}Adaptive Nuclear-Spin Sector Selection and Phase Boundary Determination}

The partition function separates as
\begin{equation}
\begin{split}
Z=\sum_{m} d(m)\,Z_m,\quad
d(m)=\binom{N}{N/2+m}\\
Z_m=\sum_{n}e^{-\beta E_n(m)},
\label{eq:Z}
\end{split}
\end{equation}
where $m$ runs from $-N/2$ to $N/2$. At low temperature $Z_m\approx e^{-\beta E_{\rm gs}(m)}$,
so the weight of sector $m$ is $W(m)=d(m)\,e^{-\beta E_{\rm gs}(m)}\equiv e^{-\beta f(m,T)}$,
with sector free energy
\begin{equation}
f(m,T)=E_{\rm gs}(m)-T\ln d(m).
\label{eq:f}
\end{equation}
The dominant sector minimizes $f$, arising from a competition between $E_{\rm gs}(m)$,
which favors large $|m|$, and $T\ln d(m)$, which peaks at $m=0$.

\textit{Ground-state energy:} Within sector $m$ the hyperfine term provides an effective
transverse field $h_{\rm eff}(m)=h_x+2Am/N$, giving the Hamiltonian
\begin{equation}
H_{\rm el}(m)=-\frac{2J_z}{N}(S_t^z)^2+2h_{\rm eff}(m)\,S_t^x.
\end{equation}
Its ground state lies in the $S_t=N/2$ sector and is well approximated by the $xz$-plane
coherent state $|\theta\rangle=\bigotimes_i(\cos\frac{\theta}{2}|\uparrow\rangle_i
+\sin\frac{\theta}{2}|\downarrow\rangle_i)$, for which
$\langle S_t^x\rangle=\frac{N}{2}\sin\theta$ and
$\langle(S_t^z)^2\rangle\approx\frac{N^2}{4}\cos^2\theta$ in the large-$N$ limit.
The variational energy per site, $-\frac{J_z}{2}\cos^2\theta+h_{\rm eff}\sin\theta$,
is minimized by $\sin\theta^*=-h_{\rm eff}/J_z$, giving
\begin{equation}
E_{\rm gs}(m)=-\frac{NJ_z}{2}-\frac{N}{2J_z}h_{\rm eff}(m)^2.
\end{equation}
Expanding $h_{\rm eff}(m)^2=(h_x+2Am/N)^2$ explicitly in $m$,
\begin{equation}
E_{\rm gs}(m)=-\frac{NJ_z}{2}-\frac{Nh_x^2}{2J_z}
-\frac{2Ah_x}{J_z}\,m-\frac{2A^2}{NJ_z}\,m^2.
\label{eq:Egs}
\end{equation}
This is a downward parabola in $m$: energy alone favors $|m|\to N/2$.

\begin{figure}[h]
\centering
\includegraphics[width=0.95\linewidth]{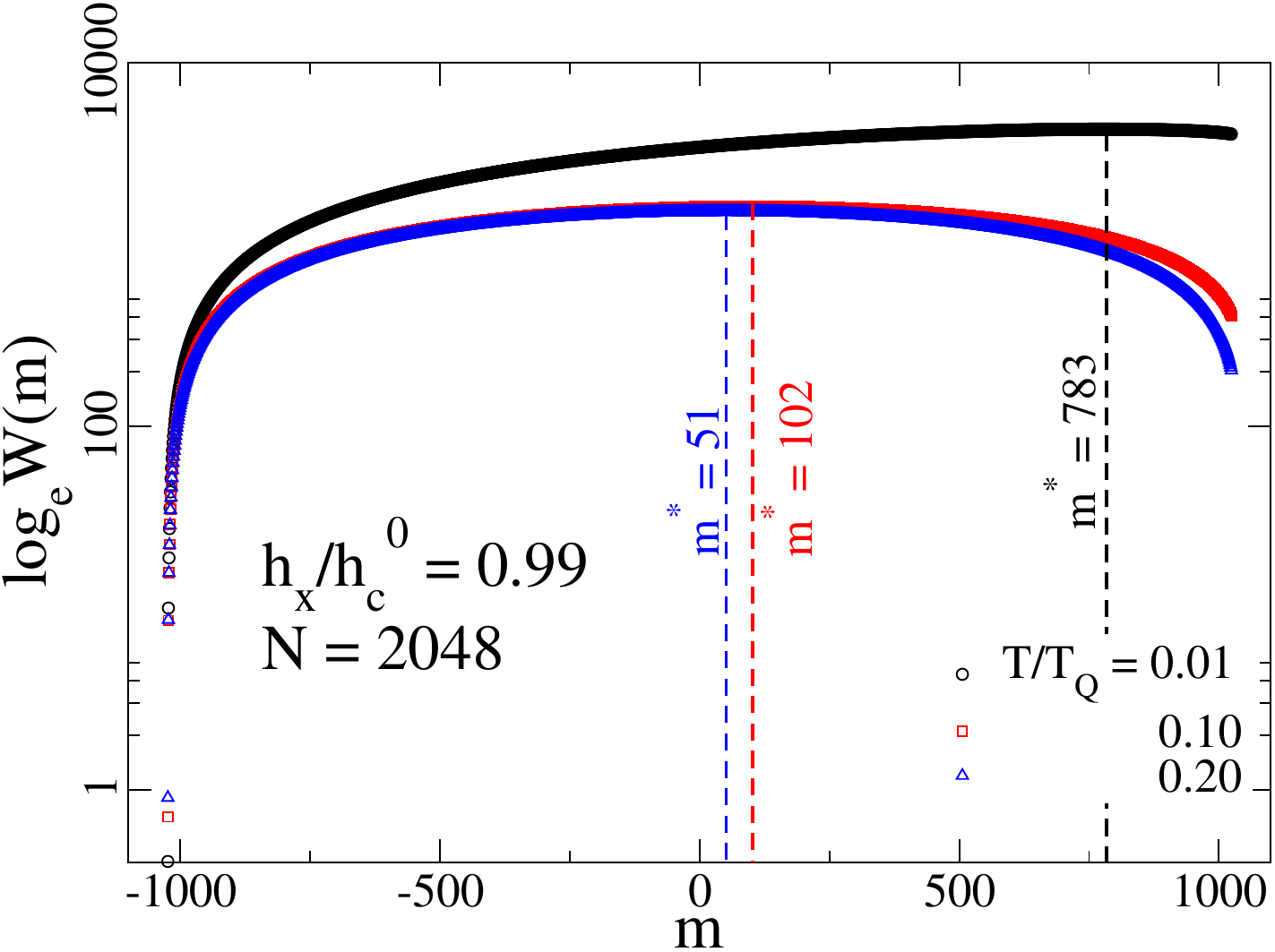}
\caption{Nuclear-spin-sector weight, shown as $\log_e W(m)$, as a function of the nuclear-spin sector $m$ for $N=2048$ and $h_x=0.99$. Results are shown for $T/T_Q=0.01$, $0.10$, and $0.20$. The vertical dashed lines indicate the most probable nuclear-spin sectors $m^*$ obtained from the Eq. \ref{eq:mstar} , with $m^*=783$, $102$, and $51$ for $T/T_Q=0.01$, $0.10$, and $0.20$, respectively. The distribution becomes progressively broader with increasing temperature, reflecting the increasing contribution of neighboring nuclear-spin sectors to the partition function.}
\label{fig:weight_vs_m}
\end{figure}

\textit{Saddle-point condition:} Using the Stirling result
$\frac{d}{dm}\ln d(m)=-2\operatorname{artanh}(2m/N)$
and setting $\partial f/\partial m=0$ with Eq. (~\eqref{eq:Egs}) yields the dominant sector
$m^*$:
\begin{equation}
\frac{2m^*}{N}=\tanh\!\left[\frac{A}{J_zT}
\!\left(h_x+\frac{2Am^*}{N}\right)\right].
\label{eq:mstar}
\end{equation}
As $T\to 0$, $m^*\to\pm N/2$ (fully polarized ground state). At $h_x=0$, a nonzero solution appears below $T_N=A^2/J_z$, recovering the nuclear ordering scale $T_N\sim A^2/J_z$; a finite $h_x$ biases $m^*$ smoothly away from zero at all $T$. Expanding the free energy about the dominant sector $m^*$ shows that the statistical weight $W(m)$ is approximately Gaussian as shown in Fig. \ref{fig:weight_vs_m} and is strongly peaked around $m^*$. Consequently, only the nuclear-spin sectors in the vicinity of $m^*$ contribute appreciably to the partition function.

\begin{figure}[t]
\centering
\includegraphics[width=0.95\columnwidth]{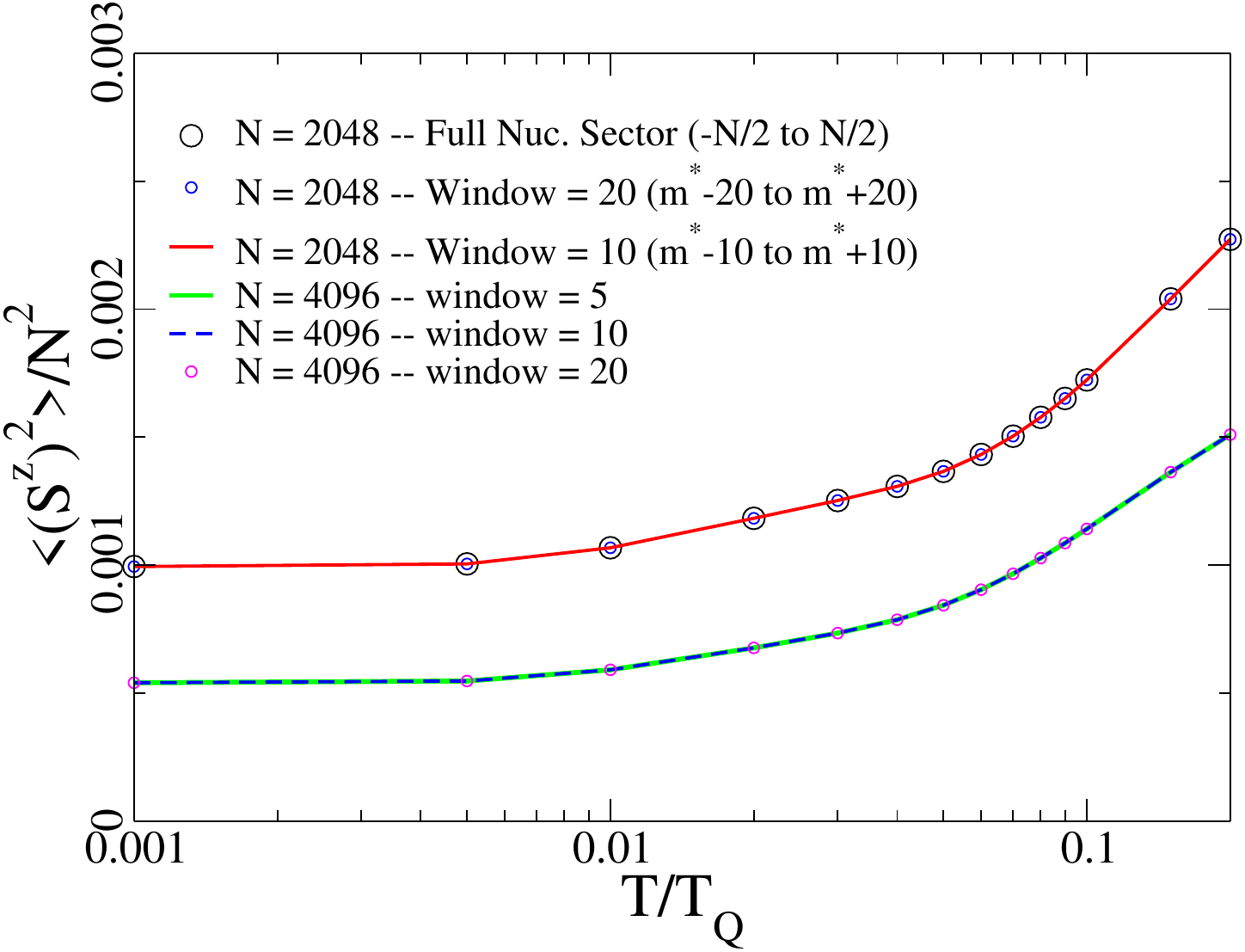}
\caption{Convergence of the electronic observable $\langle (S^z)^2\rangle/N^2$ with respect to the nuclear-spin sector window for different system sizes. For $N=2048$, the result obtained by retaining the full nuclear-spin sector, $-N/2\leq m\leq N/2$, is compared with calculations using windows of $m^*-20\leq m\leq m^*+20$ and $m^*-10\leq m\leq m^*+10$. For $N=4096$, results obtained using windows of $\pm5$, $\pm10$, and $\pm20$ around the most-weighted nuclear-spin sector $m^*$ are shown. The excellent agreement between the different window sizes demonstrates the convergence of $\langle (S^z)^2\rangle/N^2$ with a relatively small number of nuclear-spin sectors.}
\label{fig:window}
\end{figure}

\textit{Adaptive nuclear-sector selection:}  Motivated by this observation, we diagonalize the electronic pseudospin Hamiltonian only for the nuclear sectors satisfying $m^*-\Delta m \le m \le m^*+\Delta m$. In all calculations presented in this work, we choose a fixed window of $\Delta m=10$ around $m*$. We have verified that the partition function and all thermodynamic observables are well converged with respect to further enlargement of the window, demonstrating that the $\pm10$ window is sufficient for the low-temperature regime considered here as shown in Fig. \ref{fig:window}.

The present adaptive algorithm is primarily intended for investigating the low-temperature thermodynamic properties of the model, where the dominant contribution arises from a limited number of nuclear-spin sectors. To determine the phase boundary for finite hyperfine coupling, we calculate $\frac{d\langle (S^z)^2 \rangle}{dh_x}$ for system sizes $N=8192$, $10000$, $14000$, $16384$, $20000$, $24000$, and $32768$ up to $T/T_Q=0.2$. Beyond this temperature, the adaptive nuclear-sector selection algorithm becomes less reliable because the partition function receives significant contributions from a large number of nuclear-spin sectors. Consequently, retaining only a limited number of sectors is insufficient to accurately evaluate thermodynamic quantities. We have verified the convergence of the adaptive algorithm and find that it provides reliable results up to $T/T_Q=0.2$, allowing us to determine the phase boundary within this temperature range. 

For each system size, $\frac{d\langle (S^z)^2 \rangle}{dh_x}$ exhibits a
minimum at a characteristic value of $h_x$, which is identified as the
finite-size critical field $h_c(N)$. The thermodynamic-limit critical field is then obtained by extrapolating these finite-size values using
\begin{equation}
h_c(N)=h_c(\infty)+\frac{a}{N}+\frac{b}{N^2},
\end{equation} 
where $a$ and $b$ are fitting parameters. The extrapolated values yield the
blue phase boundary shown in Fig.~\ref{fig:boundary_shift_A} of the main text.

\section{\label{App:E}Finite-size scaling of the excitation gap near the critical point}
To characterize the critical behavior of the model, we study the excitation gap at the critical transverse field $h_x \approx 0.99$. 

\begin{figure}[h]
\centering
\includegraphics[width=1.0\linewidth]{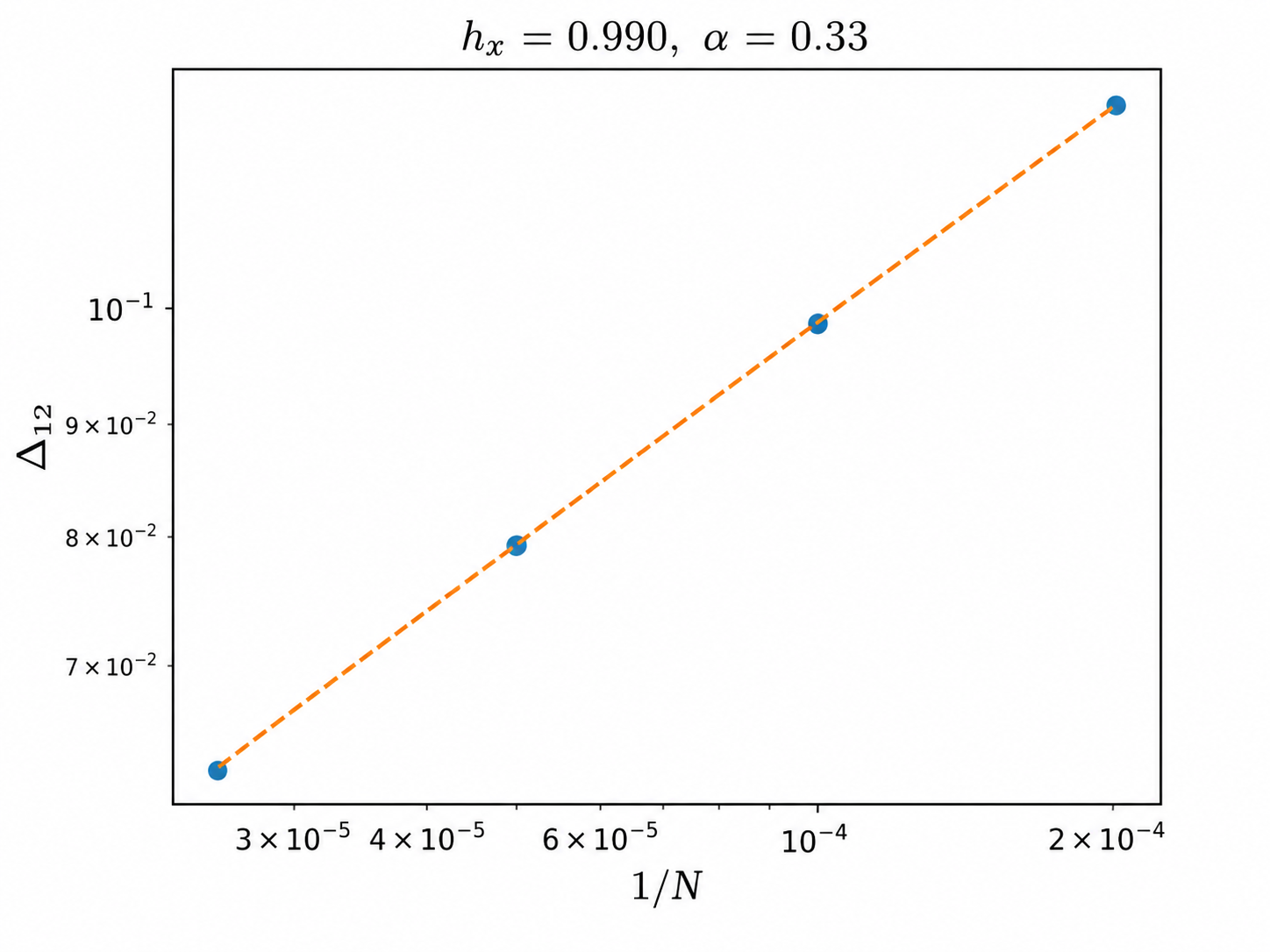}
\caption{Finite-size scaling of the excitation gap $\Delta_{12}=E_{1st}-E_{gs}$ as a function of $1/N$ at the critical point. The dashed line represents the power-law fits used to extract the scaling exponent $\alpha$. 
}
\label{fig:gap_scaling}
\end{figure}

To quantify the critical behavior, we perform a finite-size scaling analysis of the excitation gap $\Delta$ as a function of the inverse system size $1/N$. The numerical data are well described by a power-law scaling form
\begin{equation}
\Delta(N) \sim N^{-\alpha},
\end{equation}
Our numerical results show that the exponent is approximately $\alpha \approx \frac{1}{3}$ as shown in Fig. \ref{fig:gap_scaling}.


\end{document}